\documentclass[11pt,xcolor=dvipsnames]{article}
\pdfoutput=1 

\usepackage{scalerel,stackengine}
\usepackage[colorlinks,urlcolor=black,linkcolor = black,citecolor = black,]{hyperref}
\usepackage{subfigure}
\usepackage{cite}
\usepackage{ulem}
\usepackage{latexsym}
\usepackage{epsfig}
\usepackage[mathscr]{eucal}
\usepackage{amsfonts}
\usepackage{amscd}
\usepackage{array}
\usepackage{amssymb}
\usepackage{colordvi}
\usepackage[centertags]{amsmath}
\usepackage{enumerate}
\usepackage{graphicx}
\usepackage{booktabs}
\usepackage{theorem}
\usepackage[footnotesize]{caption}
\usepackage{soul}
\usepackage{slashed}
\usepackage{xcolor}
\usepackage{bbm}
\usepackage[utf8]{inputenc}
\usepackage{fancyvrb}
\usepackage{framed}
\usepackage{xspace}
\usepackage{todonotes}
\usepackage{braket}
\usepackage{listings}
\usepackage{cancel}
\usepackage{longtable}

\definecolor{codegreen}{rgb}{0,0.6,0}
\definecolor{codegray}{rgb}{0.5,0.5,0.5}
\definecolor{codepurple}{rgb}{0.58,0,0.82}
\definecolor{backcolour}{rgb}{0.95,0.95,0.92}

\lstdefinestyle{mystyle}{
    backgroundcolor=\color{backcolour},   
    commentstyle=\color{codegreen},
    keywordstyle=\color{magenta},
    numberstyle=\tiny\color{codegray},
    stringstyle=\color{codepurple},
    basicstyle=\ttfamily\footnotesize,
    breakatwhitespace=false,         
    breaklines=true,                 
    captionpos=b,                    
    keepspaces=true,                 
    numbers=left,                    
    numbersep=5pt,                  
    showspaces=false,                
    showstringspaces=false,
    showtabs=false,                  
    tabsize=2
}
\lstdefinestyle{simplestyle}{
    backgroundcolor=\color{backcolour},   
    commentstyle=\color{codegray},
    keywordstyle=\color{codegray},
    numberstyle=\tiny\color{codegray},
    stringstyle=\color{codegray},
    basicstyle=\ttfamily\footnotesize,
    breakatwhitespace=false,         
    breaklines=true,                 
    captionpos=b,                    
    keepspaces=true,                 
    numbers=left,                    
    numbersep=5pt,                  
    showspaces=false,                
    showstringspaces=false,
    showtabs=false,                  
    tabsize=2
}

\usepackage{url}

\usepackage{cleveref}
\crefname{chapter}{Chapter}{Chapter}
\crefname{section}{Sec.}{Secs.}
\crefname{table}{Tab.}{Tabs.}
\crefname{figure}{Fig.}{Figs.}
\crefname{equation}{Eq.}{Eqs.}
\crefname{appendix}{Appendix\ }{Appendix\ }
\Crefname{section}{Section}{Sections}
\Crefname{table}{Tale}{Tables}
\Crefname{figure}{Figure}{Figures}
\Crefname{equation}{Equation}{Equations}
\newcommand{\gev}{~\text{GeV}}

\newcommand{\beq}{\begin{eqnarray}} 
\newcommand{\eeq}{\end{eqnarray}} 
\newcommand{\s}{\newline \vspace*{-3.5mm}}
\newenvironment{kasten*}[1]
{
\hspace{0.05\linewidth}
\begin{minipage}{0.95\linewidth}
\setlength{\fboxsep}{10pt}
\definecolor{shadecolor}{gray}{0.9}
\definecolor{framecolor}{gray}{0}

\MakeFramed {\FrameRestore}
\subsection*{#1}
}
{
\endMakeFramed
\end{minipage}
\vspace{1em}
}
\newcommand{\download}{\url{https://github.com/phbasler/BSMPT}}

\newcommand{\home}{\$BSMPT}
\newcommand{\ewBSMPT}{\tt BSMPT v2}
\newcommand{\ii}{\mathrm{i}}
\newcommand{\VIA}{{\tt VIA}\xspace}
\newcommand{\FH}{{\tt FH \xspace}}

\newcommand{\HCW}[2]{\ensuremath{H^{\mathrm{CW}}_{#1,#2} }}
\newcommand{\NCW}[1]{\ensuremath{N^{\mathrm{CW}}_{#1} }}
\newcommand{\CW}{\text{CW}}
\newcommand{\cbrak}[1]{\left(#1\right)}
\newcommand{\primed}[1]{#1^{\prime}}
\newcommand{\vw}{v_W}

\begin{document}
\title{
  \vspace*{-3cm}
  \phantom{h} \hfill\mbox{\small KA-TP-06-2020}
  \\[1cm]
  \textbf{BSMPT v2\\
    A Tool for the Electroweak Phase Transition and \\ the Baryon
    Asymmetry of the Universe \\ in
    Extended Higgs Sectors}}

\date{}
\author{
Philipp Basler$^{1\,}$\footnote{E-mail:
  \texttt{philipp.basler@alumni.kit.edu}} ,
Margarete M\"{u}hlleitner$^{1\,}$\footnote{E-mail:
  \texttt{margarete.muehlleitner@kit.edu}} ,
Jonas M\"{u}ller$^{1\,}$\footnote{E-mail:
  \texttt{jonas.mueller@kit.edu}}
\\[9mm]
{\small\it
$^1$Institute for Theoretical Physics, Karlsruhe Institute of Technology,} \\
{\small\it 76128 Karlsruhe, Germany}}

\maketitle
\begin{abstract}
We present the {C++} code {\tt BSMPT v2} which is an extension of the
previous code {\tt BSMPT} for the calculation of
the strength of the electroweak phase transition in extended Higgs
sectors. The new version {\tt BSMPT v2} includes the features of {\tt
  BSMPT} and extends the already implemented models (the
2-Higgs-Doublet model (2HDM) in its CP-conserving and CP-violating versions
and the Next-to-2HDM) by the Complex Singlet Extension of the Standard
Model (CxSM). The major upgrade is the implementation of the
computation of the baryon asymmetry of the Universe for the CP-violating 2HDM (C2HDM), which is performed in two different
approximations. These changes and further smaller modifications
are described in this manual. Additionally, a detailed explanation of
the procedure for the implementation of new models is given, which has
also changed with respect to the previous version.
\end{abstract}
\thispagestyle{empty}
\vfill
\newpage
\setcounter{page}{1}
\section*{Program Summary}
\textbf{Program Title:} {\tt BSMPT }\\
\textbf{Licensing provisions:} GPL-3.0 License \\
\textbf{Programming Language:} \texttt{C++14}\\
\textbf{Nature of problem:} Non-minimal extended Higgs sector models
provide non-trivial vacuum structures which allow for a strong first
order electroweak phase transition. Such a phase transition is one of
the three Sakharov conditions that are required for a dynamical
generation of  the observed baryon  asymmetry of the universe (BAU)
through an electroweak phase transition. The actual calculation of the
electroweak baryogenesis requires the solution of the quantum
transport equation system describing the non-thermal equilibrium state
of the early universe during the phase transition. \texttt{BSMPT v2}
provides a numerical tool to investigate the vacuum structure of the
one-loop effective potential at finite temperature
including thermal masses, for an arbitrary
extended Higgs sector. It allows for the computation of the strength
of the electroweak phase transition for the
  implemented models. For the CP-violating 2HDM (C2HDM) also the
  generated BAU is calculated. 
For the latter task \texttt{BSMPT v2} has two
different approaches implemented for the formulation
  of the quantum transport 
equations, given by the {\tt FH} approach based on the semi-classical force
and the vacuum expectation value insertion method {\tt VIA}. 
\\ 
\textbf{Solution Method:} Numerical minimization of the one-loop
effective potenial including thermal masses, at
finite temperature with three different numerical minimizers,
\texttt{GSL}, \texttt{cmaes} and \texttt{NLopt}, in order to determine
the relevant parameters required for the phase transition dynamics at
the critical temperature and the criticial field
configuration. Furthermore, with the updated version \texttt{BSMPT v2}
it is possible to numerically solve for the C2HDM
the system of coupled differential transport equations and calculate
the BAU for this model.
\\ 
\textbf{Additional comments including restrictions and unusual
  features:} The BSM extensions are restricted to the Higgs
sectors. New gauge bosons and fermions would require an adaption in
the thermal corrections of the one-loop potential, which is not
implemented in \texttt{BSMPT v2}. 
At present, the BAU is only calculated for the C2HDM.
In the computation of the BAU the
wall velocity is an input parameter and assumed to be small, for the
VEV configuration a kink profile is assumed, the nucleation
temperature is approximated by the critical temperature. 

\section{Introduction}
One of the unsolved puzzles of particle physics is 
the observed baryon asymmetry of the Universe (BAU)
\cite{Bennett:2012zja}. Electroweak 
baryogenesis plays an important role in this context, as it
provides a mechanism to generate the BAU dynamically in the early Universe through
a strong first order electroweak phase transition (SFOEWPT) \cite{Kuzmin:1985mm,Cohen:1990it,Cohen:1993nk,Quiros:1994dr,Rubakov:1996vz,Funakubo:1996dw,Trodden:1998ym,Bernreuther:2002uj,Morrissey:2012db}
provided all three Sakharov conditions \cite{Sakharov:1967dj} are
fulfilled. Within the Standard Model (SM), in principle all three
conditions can be fulfilled, leading, however, to a phase transition
(PT) that is not of strong first order
\cite{Morrissey:2012db, Kajantie1993, Fodor:1994sj, Kajantie:1995kf, Jansen:1995yg, Kajantie:1996mn, Csikor:1998eu, Cline:2006ts}. Therefore, new physics 
extensions are required providing both additional sources of CP
violation and further scalar states that are able to trigger an
SFOEWPT. In order to answer the question if a specific
model can explain the BAU based on electroweak baryogenesis, we have
to investigate if the phase transition triggered by its Higgs
potential is of strong first order and if the baryon asymmetry
generated by the model matches the measured value of 
$5.8\cdot 10^{-10} < \eta < 6.5 \cdot 10^{-10}$ \cite{Tanabashi:2018oca}
quantitatively. The strength of the phase transition 
is defined by the quotient $\xi_c$ of the vacuum expectation value (VEV) $v_c$
at the critical temperature $T_c$, with $T_c$, defined as the temperature
where two degenerate global minima exist. Both quantities are obtained
from the loop-corrected effective Higgs potential at finite
temperature. A value of $\xi_c = v_c / T_c >1$ indicates an SFOEWPT
\cite{Quiros:1994dr,Moore:1998swa}. For the actual computation of the
BAU it is required to describe the thermal system of the early
universe.  The latter
requires the knowledge of the chemical potentials of all contributing particles in
front of the bubble wall which in turn are obtained from the solution
of the (quantum) transport equations. \s

In the following we present version 2 of the {\tt C++} code {\tt BSMPT}
(Beyond the Standard Model Phase Transitions)
\cite{Basler:2018cwe}.\footnote{Further public codes for the analysis of
the possible phases and phase transitions in the early Universe are
{\tt CosmoTransitions} \cite{Wainwright:2011kj}, {\tt PhaseTracer}
\cite{Athron:2020sbe} and {\tt BubbleProfiler}
  \cite{Athron:2019nbd}}. It extends the previous  
code for the computation of the phase transition by the calculation of
the BAU or the CP-violating 2-Higgs-Doublet Model
  (C2HDM). In detail, {\tt BSMPT v2} includes the features of the
previous version, that 
are:
\begin{itemize}
\item[$\ast$] Calculation of $v_c$, $T_c$ and $\xi_c$ from the loop-corrected
  effective potential at finite temperature
  \cite{Coleman:1973jx,Quiros:1999jp,Dolan:1973qd} including the daisy 
  resummation for the bosonic masses \cite{Carrington:1991hz}. Two different
  approximations for the treatment of the thermal masses are
  implemented, the Parwani
  \cite{Parwani:1991gq} and the
  Arnold-Espinosa method 
  \cite{Arnold:1992rz} with the latter set by default. The renormalisation of the
  potential is based on physical 'on-shell' conditions in the sense
  that the loop-corrected masses and mixing angles extracted from the
  effective potential are equal to their tree-level input values \cite{Basler:2016obg}.
\item[$\ast$] Calculation of the evolution of the VEVs as a function
  of the temperature.
\item[$\ast$] Calculation of the global minimum of the one-loop
  corrected potential at zero temperature.
\item[$\ast$] Calculation of the loop-corrected trilinear Higgs
  self-couplings in the on-shell scheme defined in \cite{Basler:2016obg}.
\end{itemize}
The first {\tt BSMPT} version includes as predefined models the CP-conserving
2-Higgs-Doublet Model (2HDM) \cite{Lee:1973iz,Branco:2011iw} and its
CP-violating version, the 
C2HDM
\cite{Branco:1985aq,Ginzburg:2002wt,Khater:2003wq,Fontes:2017zfn}, as
well as the 
Next-to-2HDM (N2HDM) 
\cite{Chen:2013jvg,Drozd:2014yla,Muhlleitner:2016mzt}.  
For applications of {\tt BSMPT} to phenomenological investigations of
these models, see \cite{Basler:2016obg, Basler:2017uxn,Basler:2019iuu}. \s

The newly implemented features in  {\tt BSMPT v2}, which will be
presented in this paper, are
\begin{itemize}
\item[$\ast$]  
The computation of the BAU for the C2HDM. It is
performed in two different approximations, the
first approach is based on the semi-classical force
  \cite{Fromme:2006cm,Fromme:2006wx,Kainulainen2001,Cline1998} yielding
  a set of fluid equations, which we refer to as \FH~approach. The second approach is  
  based on the VEV insertion approximation ({\tt VIA}) method 
  \cite{Riotto:1995hh,Riotto:1997vy}. The implementation of the 
  accordingly adapted transport equations in the latter approach
  follows \cite{Vries2018}. Since both approaches rely on the same assumption
  on the wall profile, {\tt BSMPT v2} allows for their comparison and
  discussion. The implemented wall profile is approximated by the Kink
  profile \cite{Fromme:2006wx,John:1998ip}. 
  A systematic comparison of the various approximations and their impact on non-minimal models has not been available so far. Our code therefore offers deliberately to the user the option to choose among various approximations to study their impact and get a feeling for the validity of the applied approximations.

  Note that at present in the \FH approach we assume
  small values for the wall velocity, which is an additional input parameter.
  Therefore, wall velocities with values close to the plasma
  speed should be taken with caution. An extension to arbitrary
  wall velocities as provided recently in \cite{Cline:2020jre} is left for future
  work. 
  In the \FH approach, we also implemented a second variant of the
  transport equations using plasma velocities expressed through 
  derivatives.

  So far, the computation of the BAU can be performed only for the
  C2HDM. Extensions to other models are planned for future versions of
  the code.
\item[$\ast$] Implementation of a new model, the Complex Singlet
  Extension of the SM (CxSM). 
\end{itemize}

\noindent
Further differences of {\tt BSMPT v2} with respect to the first {\tt
  BSMPT} version are
\begin{itemize}
\item[$\ast$] {\tt BSMPT} can now be exported as library.
\item[$\ast$] The interfaces of the various functions have changed.
\item[$\ast$] The code base has been improved and changed from {\tt
    C++11} to {\tt C++14}.
\end{itemize}

In this paper, we present the new features of {\tt BSMPT
  v2}, which can be downloaded at:
\begin{center}
  \download
\end{center}
We start by reviewing in \cref{sec:implmodels} the
newly implemented SM extension by a complex singlet field, the
CxSM. In \cref{sec:calcbau}, the implementation of the
calculation of the BAU for the C2HDM is presented in
detail. In \cref{sec:wallprofile} the 
applied treatment of the wall profile is discussed before presenting
in \cref{sec:fh} the \FH approach and in
\cref{sec:via} the \VIA approach for the calculation of the
BAU. \cref{sec:program} is devoted to the program description,
specifying the system requirements, the installation and usage of the
program. For completeness, we repeat the features of the
first version of {\tt BSMPT} when presenting the new modules and
functions of {\tt BSMPT v2}. 
In the computation of the baryon asymmetry we apply a number of
approximations. 
While a systematic comparison of the various approximations and their impact on non-minimal models has not been available so far, our
code has been set up such that the user has the option to choose among  various approximations which allows for the study of their impact. 
In Sec.~\ref{sec:discussion}, we present a generic discussion of the approximations
and options that are provided in order to give some guidance to the user.
The code allows for the implementation
of a new model by the user. 
Since this feature has changed in {\tt
  BSMPT v2} with
respect to the original version of the code, \cref{sec:newmodels}
explains in detail, with the help of a toy model, the
steps necessary to implement a new model in {\tt BSMPT
  v2}. Previously implemented models by the user will have to be
adapted to the new procedure in order to use {\tt BSMPT
  v2}. Sec.~\ref{sec:lib} comments on the usage of {\tt BSMPT} as
library and Sec.~\ref{sec:upgrade} on the upgrade from {\tt v1} to
{\tt v2}. We conclude in Sec.~\ref{sec:concl}. \s

The version {\tt BSMPT v2} presented in this paper can be understood as a
first proof of concept for combining our computation of the
electroweak phase transition using on-shell renormalisation with the
calculation of the BAU in two different approaches. The code will
continuously be updated in future to implement extensions of the
approximations used here and to extend it to new models. The paper
will be accompanied by a forthcoming paper \cite{EWBGPhenoPaper} applying the computation of
the BAU to a specific model including the up-to-date theoretical and
experimental constraints. 

\section{The Newly Implemented Model CxSM \label{sec:implmodels}}
The new version of {\tt BSMPT} has been extended to include an additional
model, the Complex Singlet Extension of the SM (CxSM) \cite{Barger:2008jx,Gonderinger:2012rd,Coimbra:2013qq,Jiang:2015cwa,Sannino:2015wka,Costa:2015llh,Muhlleitner:2017dkd,Chiang:2017nmu,Azevedo:2018llq}. Since the general 
calculation of the one-loop effective potential at finite temperature
does not differ from the previous version of {\tt BSMPT}, we skip the
introduction of the effective potential and refer to
\cite{Basler:2018cwe} for details, for specific model discussions we
refer to \cite{Basler:2016obg,Basler:2017uxn,Basler:2019iuu}. The Higgs
potential of the CxSM is based on the extension of the SM Higgs
potential by a complex scalar singlet field $\mathbb{S}$. The tree-level potential with a
softly broken global $U(1)$ symmetry is given by
\begin{align}
  V = & \frac{m^2}{2} \Phi^\dagger \Phi + \frac{\lambda}{4}
        \left(\Phi^\dagger \Phi\right)^2 + \frac{\delta_2}{2}
        \Phi^\dagger \Phi \vert \mathbb{S} \vert^2 + \frac{b_2}{2}
        \vert \mathbb{S} \vert^2 + \frac{d_2}{4} \vert \mathbb{S}
        \vert^4 + \left( \frac{b_1}{4} \mathbb{S}^2 + a_1 \mathbb{S} + c.c. \right) \,, 
  \label{eq:CxSMPot}
\end{align}
where 
\beq
\mathbb{S} =  \frac{1}{\sqrt{2}} \left( S + \ii A \right) 
\eeq
is a hypercharge zero scalar field. The soft-breaking terms are
written in parenthesis. After electroweak symmetry breaking the Higgs
doublet field $\Phi$ and the singlet field $\mathbb{S}$ can be written as
\begin{align}
  \Phi = & \frac{1}{\sqrt{2}} \begin{pmatrix}
    G^+ \\ v + h + \ii G^0
  \end{pmatrix} \,,        \label{eq:gaugestates1}  \\
  \mathbb{S} =    & \frac{1}{\sqrt{2}} \left( s + v_s + \ii \left( a + v_a \right) \right) \,,
  \label{eq:gaugestates} 
\end{align}
where $v \approx 246$~GeV is the SM vacuum expectation value (VEV) of
$h$ and $v_s$ and $v_a$ are the VEVs of the real and imaginary parts
of the complex singlet field, respectively. Because of the hermicity
of the potential, all parameters in \cref{eq:CxSMPot} are real, except
for $b_1$ and $a_1$. 
The complex phases can be reabsorbed through a
redefinition of the fields so that either the phase of $b_1$ or $a_1$
can be set to zero. The user can choose among the two options that
have both been implemented in {\tt BSMPT}. \s

The mass eigenstates $h_i$ $(i=1,2,3)$ are obtained from the gauge fields $h, s ,a $ in
Eqs.~(\ref{eq:gaugestates1}), (\ref{eq:gaugestates}) through a rotation 
\begin{align}
  \begin{pmatrix}
    h_1 \\ h_2 \\ h_3
  \end{pmatrix} & = R \begin{pmatrix}
    h \\ s \\ a
\end{pmatrix} \,,
\end{align}
with (${\cal M}^2$ denotes the mass matrix squared in the gauge basis)
\beq
R {\cal M}^2 R^T = \mbox{diag} (m_{h_1}^2, m_{h_2}^2, m_{h_3}^2)
\eeq
and $m_{h_1}\le m_{h_2} \le m_{h_3}$ by convention. 
The $3\times 3$ rotation matrix is parametrised in terms of three
mixing angles $\alpha_i$ $(i=1,2,3)$  
\begin{align}
  R = \begin{pmatrix} c_1 c_2        & s_1 c_2                   & s_2     \\ - (c_1 s_2
    s_3 + s_1 c_3) & c_1 c_3 - s_1 s_2 s_3     & c_2 s_3 \\ -c_1 s_2 c_3 +
    s_1 s_3        & - (c_1 s_3 + s_1 s_2 c_3) & c_2 c_3\end{pmatrix} \;,
\end{align}
where the shorthand notation $c_i \equiv \cos\alpha_i$ and
$s_i \equiv \sin\alpha_i$ has been introduced. Without loss of
generality, the interval of the mixing angles is taken as  
\begin{equation}
  -\pi/2 \le \alpha_i < \pi/2\,.
\end{equation}

The minimum conditions of the potential 
\begin{align}
  0 = & v\left(  \frac{\lambda}{4} v^2 + \frac{\delta_2}{4} \left( v_a^2 + v_s^2 \right) + \frac{m^2}{2} \right)                                                                      \\
  0 = & \mbox{Re}(a_1) \sqrt{2} - \mbox{Im}(b_1) \frac{v_a}{2} + v_s \left( \frac{d_2}{4} \left( v_s^2 + v_a^2 \right) + \frac{\delta_2}{4} v^2 + \frac{1}{2} \mbox{Re}(b_1) + \frac{1}{2} b_2 \right)  \\
  0 = & -\mbox{Im}(a_1) \sqrt{2} -\mbox{Im}(b_1) \frac{v_s}{2} + v_a \left( \frac{d_2}{4} (v_s^2 + v_a^2) + \frac{\delta_2}{4} v^2 - \frac{1}{2} \mbox{Re}(b_1) + \frac{b_2}{2} \right) 
\end{align}
are used to express the parameters of the potential in terms of the
input parameters. These depend on the choice of the VEV configuration
of the singlet field, {\it cf.}~App.~\ref{app:tadpole}.\s

The counterterm potential of the CxSM reads
\begin{align}
  V^{\mathrm{CT}} = & \frac{\delta m^2}{2} \Phi^\dagger \Phi +
                      \frac{\delta \lambda}{4} \left(\Phi^\dagger
                      \Phi\right)^2 + \frac{\delta \delta_2}{2}
                      \Phi^\dagger \Phi \vert \mathbb{S} \vert^2 +
                      \frac{\delta b_2}{2} \vert \mathbb{S} \vert^2 +
                      \frac{\delta d_2}{4} \vert \mathbb{S} \vert^4 \\\nonumber
                    &+  \left( \frac{1}{2} \delta\mbox{Re}(b_1) \left(
                      (\mbox{Re} \mathbb{S})^2 -(\mbox{Im}
                      \mathbb{S})^2 \right) - \mbox{Re} \mathbb{S}\,
                      \mbox{Im} \mathbb{S} \,\delta\mbox{Im} b_1 +2
                      \delta a_1 \,\mbox{Re} \mathbb{S}  - 2 \delta
                      a_1 \,\mbox{Im} \mathbb{S} \right)  \\ \nonumber
                    & + \delta T_h (h + \omega_h) + \delta T_s \left(
                      s+\omega_s\right) + \delta T_a \left( a +
                      \omega_a\right)  \,, 
\end{align}
with the tadpole counterterms $\delta T_i$ for each field with a
non-zero VEV. We use the symbol $\omega$ to generically denote
VEVs including loop and non-zero temperature effects in the
potential. The renormalisation conditions presented in
\cite{Basler:2016obg,Basler:2017uxn,Basler:2018cwe,Basler:2019iuu} allow
to use the loop-corrected masses and mixing angles as direct inputs of the
parameter scan
since these conditions require the loop-corrected masses
  derived from the effective potential to be equal to their tree-level
  values. In this sense we call them 'on-shell' although this
  expression should be taken with caution here, since the
  loop-corrected masses extracted from the effective potential are
  evaluated at vanishing external momentum, whereas on-shell
  conditions in the proper sense demand the renormalised mass to be
  equal to the tree-level one at momentum squared equal to the mass
  squared of the particle. While the zero-momentum approximation
  used in the 'on-shell' definition deteriorates with 
  rising particle mass the loop corrections to heavy particles are
  in general small\footnote{See for example Ref.~\cite{Slavich:2020zjv} for an
    overview of the computation of mass
    corrections in the effective potential approach and explicit
    diagrammatic calculations in supersymmetric models.} on the
other hand.  Therefore, we expect the impact of the missing pieces to
be small. \s

In the following we list the counterterms for the
various possible VEV configurations, where we use the shorthand notations 
\begin{equation}
  \NCW{\phi_i}=\frac{\partial V^{\text{CW}}}{\partial\phi_i}\,,\quad\HCW{\phi_i}{\phi_j} = \frac{\partial^2 V^{\text{CW}}}{\partial \phi_i\partial\phi_j}\,,
\end{equation}
for the first and second derivatives of the Coleman-Weinberg
potential. The parameters $t_i \in\mathbb{R}$ $(i=1,2,3,4)$ that appear
in the following counterterms can be chosen freely. For
simplicity, in the implementation in {\tt BSMPT}, they are set to
zero. 
\paragraph*{Case $v_a \neq 0$ and $v_s \neq 0$:}
\begin{subequations}
\begin{align}
  \delta\lambda =  & \frac{2}{v^2} \left( \HCW{G_0}{G_0} - \HCW{h}{h} \right)  \\
  \delta m^2 =     & \frac{v_a}{v} \HCW{h}{a} -3 \HCW{G_0}{G_0} + \HCW{h}{h} + \frac{v_s^2}{v v_a} \HCW{h}{a} \\
  \delta\delta_2 = & -\frac{2}{v_a v} \HCW{h}{a} \\
  \delta b_2 =     & \HCW{h}{a} \frac{v}{v_a} - \HCW{s}{s} + 2\frac{v_a}{v_s} \HCW{s}{a} - \HCW{a}{a} - t_1 \left( \frac{v_a}{v_s} + \frac{v_s}{v_a} \right) \\
  \delta d_2 =     & -\frac{2}{v_av_s} \HCW{s}{a} + \frac{t_1}{v_a v_s}  \\
  \delta \mbox{Re} b_1 = & \HCW{a}{a} - \HCW{s}{s} + \HCW{s}{a} \left( \frac{v_s}{v_a} - \frac{v_a}{v_s} \right) + t_1 \left( \frac{v_a}{2v_s} - \frac{v_s}{2v_a} \right) \\
  \delta \mbox{Im} b_1 = & t_1 \\
  \delta \mbox{Re} a_1 = & \frac{1}{\sqrt{2}} \left( \HCW{s}{s} v_s - \HCW{s}{a} \frac{v_s^2}{v_a} - \NCW{s} \right) - \frac{1}{\sqrt{2}} t_2 + \frac{\sqrt{2}}{4} t_1 \left( v_a + \frac{v_s^2}{v_a} \right)  \\
  \delta \mbox{Im} a_1 = & \frac{1}{\sqrt{2}} \left( \HCW{s}{a} \frac{v_a^2}{v_s} - \HCW{a}{a} v_a + \NCW{a} \right) + \frac{1}{\sqrt{2}} t_3 - \frac{\sqrt{2}}{4} t_1 \left( \frac{v_a^2}{v_s} +  v_s \right) \\
  \delta T_h =     & \HCW{G_0}{G_0} v - \NCW{h}  \\
  \delta T_s =     & t_2 \\
  \delta T_a =     & t_3 \,.
\end{align}  
\end{subequations}

\paragraph*{Case $v_s \neq 0$ and $v_a = 0$:}
\begin{subequations}  
  \begin{align}
    \delta m^2 =      & \HCW{h}{h} - 3 \HCW{G_0}{G_0} +\frac{v_s}{v} \HCW{h}{s}                                                                 \\
    \delta\lambda =   & \frac{2}{v^2} \left( \HCW{G_0}{G_0} - \HCW{h}{h} \right)                                                                \\
    \delta \delta_2 = & -\frac{2}{v v_s} \HCW{h}{s}                                                                                             \\
    \delta b_2 =      & \HCW{s}{s} - \HCW{a}{a} + \frac{v}{v_s} \HCW{h}{s} - \frac{2}{vs} \NCW{s} -\frac{2\sqrt{2}}{v_s} t_1 - \frac{2}{v_s} t_2 \\
    \delta d_2 =      & \frac{1}{v_s^3} \left( -2 \HCW{s}{s} v_s + 2\sqrt{2} t_1 + 2\NCW{s} + 2 t_2 \right)                          \\
    \delta \mbox{Re} b_1 =  & \HCW{a}{a} -\frac{\sqrt{2}}{v_s} t_1 + \frac{1}{v_s} \NCW{s}  - \frac{t_2}{v_s}                                      \\
    \delta \mbox{Im} b_1 =  & 2 \HCW{s}{a}                                                                                                          \\
    \delta \mbox{Re} a_1 =  & t_1                                                                                                                 \\
    \delta \mbox{Im} a_1 =  & \frac{1}{\sqrt{2}} \left( t_3 + \NCW{a} - \HCW{s}{a} v_s \right)                                                     \\
    \delta T_h =      & \HCW{G_0}{G_0} v - \NCW{h}                                                                                           \\
    \delta T_s =      & t_2                                                                                                                  \\
    \delta T_a =      & t_3 \,.
  \end{align}
\end{subequations}
\paragraph*{Case $v_a \neq 0$ and $v_s = 0$:}
\begin{subequations}  
  \begin{align}
    \delta m^2 =      & \HCW{h}{h} - 3 \HCW{G_0}{G_0} + \frac{v_a}{v} \HCW{h}{a}                                                                 \\
    \delta \lambda =  & \frac{2}{v^2} \left( \HCW{G_0}{G_0} - \HCW{h}{h} \right)                                                            \\
    \delta \delta_2 = & -\frac{2}{v v_a} \HCW{h}{a}                                                                                           \\
    \delta b_2 =      & \frac{2\sqrt{2}}{v_a} t_1 -\frac{2}{v_a} t_3 \HCW{a}{a} - \HCW{s}{s} + \frac{v}{v_a} \HCW{h}{a} - \frac{2}{v_a} \NCW{a} \\
    \delta d_2 =      & \frac{2}{v_a^3} \left( t_3 - \sqrt{2} t_1 \right) + \frac{2\NCW{a} - 2\HCW{a}{a} v_a}{v_a^3}                       \\
    \delta \mbox{Re} b_1 =  & \frac{t_3 - \sqrt{2}t_1}{v_a} + \frac{\NCW{a}}{v_a} - \HCW{s}{s}                                                     \\
    \delta \mbox{Im} b_1 =  & 2 \HCW{s}{a}                                                                                                          \\
    \delta \mbox{Re} a_1 =  & - \frac{1}{\sqrt{2}} \left( t_2 + \NCW{s} - \HCW{s}{a} v_a \right)                                                       \\
    \delta \mbox{Im} a_1 =  & t_1                                                                                                               \\
    \delta T_h =      & \HCW{G_0}{G_0} v - \NCW{h}                                                                                             \\
    \delta T_s =      & t_2                                                                                                                   \\
    \delta T_a =      & t_3 \,.
  \end{align}
\end{subequations}
\paragraph*{Case $v_a = v_s = 0$:}
\begin{subequations}
  \begin{align}
    \delta m^2 =      & \HCW{h}{h} - 3 \HCW{G_0}{G_0}                     \\
    \delta \lambda = & \frac{2}{v^2} \left( \HCW{G_0}{G_0} - \HCW{h}{h} \right)     \\
    \delta \delta_2 = & t_1                                               \\
    \delta b_2 =      & -\frac{v^2}{2} t_1 - \HCW{s}{s} - \HCW{a}{a}    \\
    \delta d_2 =      & t_2                                               \\
    \delta \mbox{Re} b_1 =  & \HCW{a}{a} - \HCW{s}{s}                         \\
    \delta \mbox{Im} b_1 =  & 2 \HCW{s}{a}                                   \\
    \delta \mbox{Re} a_1 =  & - \frac{1}{\sqrt{2}} \left( t_3 + \NCW{s} \right) \\
    \delta \mbox{Im} a_1 =  & \frac{1}{\sqrt{2}} \left( t_4 + \NCW{a} \right) \\
    \delta T_h =      & \HCW{G_0}{G_0} v - \NCW{h}                    \\
    \delta T_s =      & t_3                                           \\
    \delta T_a =      & t_4 \,.
  \end{align}
\end{subequations}
Note that we treat the complex singlet field with respect to the
minimisation in all cases in the same way regardless of the VEV
configuration, meaning that the singlet fields $s ,a $ in
Eq.~(\ref{eq:gaugestates}) are allowed to
evolve a non-zero VEV at finite 
temperature even when the zero temperature values are chosen to be
zero. In this way it is possible to investigate the stability of the
Dark Matter candidate at non-zero temperatures. 

\section{Calculation of the Baryon Asymmetry of the Universe \label{sec:calcbau}}
In this section we will briefly document the approaches
implemented in {\tt BSMPT v2} used for the calculation of BAU. So
far, two non-local approaches are 
implemented. These are the semi-classical force approach \cite{Fromme:2006cm,Fromme:2006wx,Kainulainen2001,Cline1998}, based on a WKB approximation of the (quantum)
transport equations, and the competing VEV-insertion approach
\cite{Riotto:1995hh,Riotto:1997vy}. Both of them rely on the same assumptions
concerning the wall profile, so that the results
provided by {\tt BSMPT} for the two different implementations allow for
their comparison and discussion.
We will start by reviewing the
calculation of the bubble wall profile, followed by the calculation of
the wall thickness. We will then list the formulas required in the two approaches for the
transport equations.   

\subsection{Wall Profile \label{sec:wallprofile}}
Both approaches assume a planar bubble wall. The thus omitted
curvature terms in the wall profile simplify the derivation of the
transport equations significantly. The dimensionality of the problem
reduces from $3+1$ to an effective $1+1$ problem, with only one space
coordinate $z$. The space coordinate will be referred to as bubble
wall distance in the following. In a first approach,
  the wall profile is approximated by the Kink profile
\cite{Fromme:2006wx},  
\begin{equation}
  f(z) = \frac{f_0}{2} \cbrak{1-\tanh\frac{z}{L_W}}\,,
  \label{eq:Kink}
\end{equation}
where $f_0$ indicates the value of the VEV inside the broken phase and
$L_W$ describes the wall thickness. 
While this is not an optimal approach, it is an
  approach that is frequently used in the literature. We varied the
  wall thickness for specific benchmark points to investigate the
  impact on the BAU. It turned out that the impact on the $\eta$
  value was in the percentage region. We plan to investigate the
  true bounce solution in future, which would allow us to estimate the
  true error.
\s

The critical temperature $T_c$ is defined by the
coexistence of two degenerate minima of the potential, the broken
non-zero VEV and the symmetric vanishing one. The wall thickness $L_W$ can be related
to the barrier height $V_b$ between both minima as \cite{Fromme:2006wx}
\begin{equation}
  L_W = \frac{v_c}{\sqrt{8 V_b}}\,,
\end{equation}
with the critical VEV $v_c$ at $T_c$ given by the VEV of the broken minimum. In fact, the
actual nucleation of the bubbles happens at the nucleation temperature
$T_N<T_c$. Since \texttt{BSMPT v2} cannot estimate the nucleation temperature, the wall thickness and thus the bubble wall profile are determined at the critical temperature. \s

The calculation of the barrier height between both minima is
implemented as follows: A straight line between both
minima in the VEV vector space $\lbrace \vec{\omega}\rbrace$ is taken
as a first guess for the tunnel path, parametrised with
$t\in\left[0,1\right]$ as
\begin{equation}
  \Gamma : \vec{\omega}(t) = \vec{\omega}_s + t \vec{n}\,,\quad\text{with}\quad \vec{n}=\vec{\omega}_b - \vec{\omega}_s\,.
\end{equation}
The broken VEV configuration $\vec{\omega}_b$ is evaluated by the {\tt
  BSMPT} routine. The complex phase of the symmetric VEV configuration is arbitrary due to the vanishing VEV. However, we want to define the symmetric phase such that we have a smooth function if an infinitesimal step is taken along the tunnel path. For that reason, we determine the global minimum in the orthogonal plane
  \begin{equation}
    \vec{\omega} \in \lbrace \vec{x} \big\vert \cbrak{\vec{x}-\cbrak{\vec{\omega}_s+\varepsilon \vec{n}}}\cdot \vec{n} = 0  \rbrace\,,
    \label{approach}
  \end{equation}
  where already an infinitesimal step in $\varepsilon \vec{n}$ is taken. Furthermore, since we define the critical temperature to be the lower bound of the bisection interval, we have to ensure that the numerical minimization finds the symmetric global minimum. This is ensured by shifting the temperature by $T_c+\delta T$ and so the temperature is above the critical temperature and the symmetric vacuum is the only global minimum of the potential. In this way we can ensure that the numerical minimization finds unambiguously the symmetric vacuum state $\vec{\tilde{\omega}}_s$ with a unique phase. Note that $\vec{\tilde{\omega}}_s$ corresponds to the found global minimum in the orthogonal plane defined in \cref{approach}. The default values of the numerical parameters are chosen to be 
  \begin{equation}
    \varepsilon = 10^{-2}\,, \qquad \delta T = 1~\mathrm{GeV}\,.
  \end{equation}
  It was explicitly checked that the impact of the chosen parameters
  has a negligible impact on the BAU prediction.
Taking the straight line in the VEV vector space $\lbrace
\vec{\omega}\rbrace$ as a first guess allows us to go from one minimum to
the other with a certain stepsize 
\begin{equation}
  \vec{\omega}\rightarrow \vec{\omega}^{(n)}\equiv\vec{\omega}(t_n)\,,
\end{equation}
where $t_n$ determines the position after the $n$-th step. The
orthogonal plane to the straight line in a given step point is defined
by  
\begin{equation}
  \vec{\omega} \in \left\{ \vec{x} \left| \cbrak{\vec{x}-\vec{\omega}^{(n)}}\cdot \vec{n}=0 \right. \right\}\,,
  \label{eq:plane_def}
\end{equation}
and allows to find the global minimum fulfilling
\cref{eq:plane_def}. Successively, the global minima in the orthogonal
planes are determined. These points create a grid that approximates
the true tunnel path between the two degenerate minima. With the help
of a cubic-spline path $\gamma$ approximating the grid, the maximum of
the effective one-loop potential at finite temperature along the
approximated true tunnel path can be determined. The resulting barrier
height is then given by  
\begin{equation}
  V_b = \max_{\vec{\omega}\in\gamma}\cbrak{
    V_{\text{eff}}\cbrak{\vec{\omega}}} -
  V_{\text{eff}}(\vec{\omega}_b) \,. 
\end{equation}\\
With the determination of the barrier height and the critical VEV
$v_c$ it is possible to parametrise the VEV configuration as a
function of the bubble distance $z$ by using \cref{eq:Kink}. 
In CP-violating models the complex VEV configuration
induces complex masses for 
fermions with Yukawa-like interactions where we parametrise the
complex fermion masses as ($i$ denotes the fermion species)
\begin{equation}
  m_i(z) = \left| m_i(z)\right| \exp(\ii \theta(z))\,.
\end{equation}
In {\tt \ewBSMPT} the absolute value $\left|m_i\right|$ and the phase
factor $\theta$ are determined numerically. The absolute value is
determined by evaluating the mass matrix with the given VEV
configuration at the wall distance $z$. The phase factor is determined
by first evaluating the phase factor in the broken and symmetric
minimum\footnote{Note that we use $\vec{\tilde{\omega}}_s$ for the symmetric vacuum configuration to ensure a unique phase of the symmetric vacuum.}, referred to as $\theta_{\text{brk}}$ and
$\theta_{\text{sym}}$, respectively, and then using
the analogous of
\cref{eq:Kink}, yielding
\begin{equation}
  \theta(z) =
  \cbrak{\theta_{brk}-\frac{\theta_{brk}-\theta_{sym}}{2}\cbrak{1+\tanh
      \frac{z}{L_W}}}\,. 
\end{equation}

\subsection{Semi-Classical Force Approach \label{sec:fh}}
In the following section we give all necessary formulas for the
numerical implementation of the semi-classical force
approach, which we denote by {\tt FH}. We do
not comment on the actual derivation of the formulas and will refer to
the corresponding literature \cite{Fromme:2006wx,Fromme:2006cm,Kainulainen2001}. For a phenomenological discussion of our
implementation we refer to \cite{EWBGPhenoPaper}. The BAU is
calculated in a two-step approach, where in the first step the quantum
transport equations are solved to obtain the left-handed fermion
excess in front of the bubble wall. In the second step this
left-handed fermion excess is translated to the actual
baryon-asymmetry through an electroweak sphaleron transition. \s

The left-handed fermion excess (given in terms of the
second-order CP-odd 
chemical potentials $\mu_{x,2}$) triggering the
electroweak sphaleron transition is given by  
\begin{align}
  \mu_{B_L} =& \frac{1}{2} \left(1+4K_{1,t}\right) \mu_{t,2} +
               \frac{1}{2} \left(1+4K_{1,b}\right) \mu_{b,2} -
               2K_{1,t} \, \mu_{t^c,2} \,, 
  \label{eq:FH_MUBL}
\end{align}
where $t (t^c)$ corresponds to the left-handed (charge conjugated
right-handed) top quark and $b$ to the bottom quark.  
The thermal transport coefficients, required here and
  in the following, are defined as
\begin{subequations}
  \begin{align}
      K_{1,i} =& - \Braket{\frac{p_z^2}{E_0} \partial_E^2 f_{i,0} } \,, \\
      K_{2,i} =& \Braket{\frac{\partial_E^2 f_{i,0}}{2E_0} } \,,\\
      K_{4,i} =& \Braket{\frac{p_z^2}{E_0^2} \partial_E f_{i,0} }\,, \\
      \tilde{K}_{5,i} =& \left[ \frac{p_z^2}{E_0} \partial_E f_{i,0} \right]\,, \\
      \tilde{K}_{6,i} =& \left[ \frac{E_0^2-p_z^2}{2E_0^3} \partial_E f_{i,0} \right]\,, \\
      K_{8,i} =& \Braket{\frac{\vert p_z\vert \partial_E f_{i,0}}{2E_0^2 E_{0z}} } \,,\\
      K_{9,i} =& \Braket{\frac{\vert p_z\vert}{4E_0^3 E_{0z}} \left(
                 \frac{\partial_E f_{i,0}}{E_0} - \partial_E^2 f_{i,0}
                 \right) }  \;,
\end{align}
\label{Eq:Kfactors}
\end{subequations}
where 
\begin{eqnarray}
E_0 (z) &=& \sqrt{p_x^2+p_y^2+p_z^2+m(z)^2}\,,\\
E_{0z} &=& \sqrt{p_z^2+m(z)^2}\,,  
\end{eqnarray}
are the energies of the quasi-particles in front of the bubble wall.
Here $\partial_E$ denotes the partial derivative with respect to the
energy of the distribution function. The expectation values are given by 
\begin{align}
  \Braket{X} = \frac{\int \mathrm{d}^3p X(p)}{\int \mathrm{d}^3p \partial_E f_{0+}(m=0)}  \,,\quad
  \left[X\right] = \frac{\int \mathrm{d}^3p X(p)}{\int \mathrm{d}^3p f_{i,0,\vw}} = \frac{\int \mathrm{d}^3p X(p)}{\int \mathrm{d}^3p f_{i,0}}  \,,
\end{align}
and the distribution functions are given by
\begin{align}
    f_{i,0} = \left. f_{i}\right|_{\mu_i = 0, \delta f_i = 0, \vw = 0} \,,\quad
    f_{0+} = \left. f_{i}\right|_{i=\mathrm{fermion},\mu_i=0,\delta f_i = 0, \vw = 0}\,, \quad
    f_{i,0,v_W} =  f_{i,0} + \vw p_z \partial_{E_0}  f_{i,0}\,.
\end{align}
Note that we do not use the complete dependence on the
wall velocity $v_W$ in the thermal coefficients induced by the
distribution function ($+$ refers to fermions, $-$ to
  bosons, $\beta=1/T$)
\begin{align}
  f_i =& \frac{1}{\exp\cbrak{\beta\left[\gamma_W \cbrak{E_0+v_W p_z}-\mu_i\right]}\pm 1} +\delta f_i
\end{align}
with
\begin{align}
 \gamma_W =& \frac{1}{\sqrt{1-v_W^2}} \,,
\end{align}
and the small perturbation $\delta f_i$.
Instead, the additional assumption of small wall
velocities ($v_W \ll1$) is used to expand the thermal coefficients,
simplifying the transport equations further. In this way, it is not
possible to discuss the behaviour for arbitrary wall velocities, and wall
velocities with values close to the plasma speed should be taken with caution. Recently,
Ref.~\cite{Cline:2020jre} extended the semi-classical 
force approach for arbitrary wall velocities. Furthermore, Ref.~\cite{Konstandin:2014zta} also improved the fluid approach to relativistic wall velocities.\s

 The thermal coefficients are evaluated in an equidistant grid in
 $\lbrace m^2(z),T\rbrace$ which is
 called through a bi-cubic spline in {\tt\ewBSMPT}. Thereby it is
 possible to keep the $z$ dependence in the thermal coefficients
 without significant increase of run-time. 
More specifically, we sampled $m^2$ from
  $0~\mathrm{GeV}^2$ to $5~\mathrm{GeV}^2$ with a step size of
  $10^{-3}$ and afterwards up to $200^2~\mathrm{GeV}^2$ with a step
  size of $5~\mathrm{GeV}^2$. The temperature steps are sampled from
  $10~\mathrm{GeV}$ up $250~\mathrm{GeV}$ with a step size of
  $2~\mathrm{GeV}$. We tested several different step sizes in the grid
  and observed that the overall result of the BAU does not show a
  significant dependence. The effect was in the sub-percentage region
  at most. The choice was a compromise between the header file size
  due to the larger grid and the required
  precision.\footnote{Information on the grid can also be found at the
  webpage in
  \url{https://phbasler.github.io/BSMPT/Kfunctions__grid_8h.html}.} 
\s

In the thermal plasma in front of the bubble wall, the full particle content has to be
considered in the transport equations. The dominant
contribution to the production of the baryon asymmetry is given by the top
quark, however, because of its large mass, or rather its strong
Yukawa interaction with the thermal plasma. It is the
  (large) amount of CP violation in the quark sector that dominantly drives
the generation of the left-handed excess. \s 

The transport equations are derived for a single 
  Dirac fermion with a complex mass term and include Yukawa
interactions, strong sphaleron interactions and $W$-boson scattering. The
transport equations read \cite{Fromme:2006cm,Fromme:2006wx} 
\begin{subequations}
  \begin{align}
      0 =& 3 \vw K_{1,t} \left( \partial_z \mu_{t,2} \right) + 3\vw K_{2,t} \left( \partial_z m_t^2 \right) \mu_{t,2} + 3 \left( \partial_z u_{t,2} \right) \notag 
      \\ &- 3\Gamma_y \left(\mu_{t,2} + \mu_{t^c,2} + \mu_{h,2} \right) - 6\Gamma_M \left( \mu_{t,2} + \mu_{t^c,2} \right) - 3\Gamma_W \left( \mu_{t,2} - \mu_{b,2} \right) \notag 
      \\ &- 3\Gamma_{ss} \left[ \left(1+9 K_{1,t} \right) \mu_{t,2} + \left(1+9 K_{1,b} \right) \mu_{b,2} + \left(1-9 K_{1,t} \right) \mu_{t^c,2} \right] \label{Eq:TransportEquations:mut} \,,\\
      0 =& 3\vw K_{1,b} \left(\partial_z \mu_{b,2}\right) + 3 \left(\partial_z u_{b,2} \right) - 3\Gamma_y \left( \mu_{b,2} + \mu_{t^c,2} + \mu_{h,2} \right) - 3\Gamma_W \left( \mu_{b,2} - \mu_{t,2} \right) \notag \label{Eq:TransportEquations:mub} \\
          &- 3\Gamma_{ss} \left[ \left( 1 + 9K_{1,t}\right) \mu_{t,2} + (1+9K_{1,b}) \mu_{b,2} + (1-9K_{1,t}) \mu_{t^c,2} \right] \,,\\
      0=&  3 \vw K_{1,t} \left( \partial_z \mu_{t^c,2} \right)  + 3\vw K_{2,t} \left( \partial_z m_t^2 \right)  \mu_{t^c,2} + 3 \left( \partial_z u_{t^c,2} \right) \notag \\
          &- 3\Gamma_y \left(\mu_{t,2} + \mu_{b,2} + 2\mu_{t^c,2} + 2\mu_{h,2} \right) - 6\Gamma_M \left( \mu_{t,2} + \mu_{t^c,2} \right) \notag \\
          &- 3\Gamma_{ss} \left[ \left( 1+9 K_{1,t}\right) \mu_{t,2} + \left(1+9K_{1,b}\right) \mu_{b,2} + \left(1-9K_{1,t}\right) \mu_{t^c,2} \right] \label{Eq:TransportEquations:mutc} \,,\\
      0 =& 4\vw K_{1,h} \left( \partial_z \mu_{h,2}\right) + 4\left( \partial_z u_{h,2}\right) - 3\Gamma_y \left( \mu_{t,2} + \mu_{b,2} + 2\mu_{t^c,2} + 2\mu_{h,2} \right) - 4\Gamma_h \mu_{h,2} \label{Eq:TransportEquations:muh} \,,\\
      S_t =& -3K_{4,t} \left( \partial_z \mu_{t,2}\right) + 3\vw \tilde{K}_{5,t} \left( \partial_z u_{t,2}\right) + 3\vw \tilde{K}_{6,t} \left( \partial_z m_t^2 \right) u_{t,2} + 3\Gamma_t^\mathrm{tot} u_{t,2} \label{Eq:TransportEquations:ut} \,,\\
      0 =& -3K_{4,b} \left( \partial_z \mu_{b,2} \right) + 3\vw \tilde{K}_{5,b} \left(\partial_z u_{b,2}\right) + 3\Gamma_b^\mathrm{tot} u_{b,2} \label{Eq:TransportEquations:ub} \,,\\
      S_t = & -3K_{4,t} \left( \partial_z \mu_{t^c,2}\right) + 3\vw \tilde{K}_{5,t} \left( \partial u_{t^c,2}\right) + 3\vw \tilde{K}_{6,t} \left( \partial_z m_t^2\right) u_{t^c,2} + 3\Gamma_t^\mathrm{tot} u_{t^c,2} \label{Eq:TransportEquations:utc} \,,\\
      0 = & -4K_{4,h} \left( \partial_z \mu_{h,2} \right) + 4\vw \tilde{K}_{5,h} \left( \partial_z u_{h,2} \right) + 4\Gamma_h^\mathrm{tot} u_{h,2} \label{Eq:TransportEquations:uh} \,.
  \end{align}
\label{Eq:TransportEquations}
\end{subequations}
The second-order CP-odd chemical potentials $\mu_{i,2}$ ($i=t,b,t^c,h$)
describe the particle density excess of the left-handed top and 
bottom quarks, the right-handed top quark and the Higgs boson. The plasma
velocities $u_i$ are defined as in Ref.~\cite{Fromme:2006cm}. The
source term $S_t$ reads
\begin{align}
  S_t =& -\vw K_{8,t} \partial_z \left( m_t^2 \partial_z \theta \right) + \vw K_{9,t} \left( \partial_z \theta \right) m_t^2  \left( \partial_z m_t^2\right) \label{Eq:TransportEquations:Source} \,.
\end{align}
The Yukawa interaction rate for the top quark, $\Gamma_y$, and the
strong sphaleron rate $\Gamma_{ss}$ are given by \cite{Fromme:2006cm}
\begin{equation}
  \Gamma_y = 4.2\cdot 10^{-3} T_c\,,\quad \Gamma_{ss} = 4.9 \cdot 10^{-4} T_c\,,
\end{equation}
and the Higgs number violating rate is given by
\begin{equation}
  \Gamma_h = \frac{m^2_W(z,T_c)}{50 T_c}\,.
\end{equation}
The $W$-boson mass $m_W$ is determined at given $z$ and for the critical
temperature $T_c$ from the numerical evaluation of the gauge boson
mass matrix. The top-helicity flipping rate is implemented as
\begin{equation}
  \Gamma_M = \frac{m^2_t(z,T_c)}{63 T_c}\,.
\end{equation}
The total quark and Higgs interaction rates can be
determined with the help of their respective diffusion constants, 
\begin{align}
  \Gamma_t^{\text{tot}}  = \frac{K_{4,t}}{K_{1,t} D_t}\,,\quad  \Gamma_b^{\text{tot}} = \frac{K_{4,b}}{K_{1,b} D_b}\,,\quad\Gamma_h^{\text{tot}}  =  \frac{K_{4,h}}{K_{1,h} D_h}\,,
\end{align} 
with the diffusion constants for the top and bottom quarks ($q=t,b$) and the Higgs boson given by \cite{Joyce1994,Cline2000a} 
\begin{equation}
  D_q = \frac{6}{T_c} \quad\text{and}\quad D_h=\frac{20}{T_c}\,.
\end{equation}
The $W$-scattering rate is approximated by the total Higgs interaction
rate $\Gamma_W = \Gamma_h^{\text{tot}}$. It should be emphasized that all
interaction rates are calculated in the plasma frame. 
The transport equations in \cref{Eq:TransportEquations} are
implemented in {\tt \ewBSMPT} and solved with the help of the {\tt
  C++}
numerical library {\tt boost} \cite{Boost1.66}, where all
chemical potentials and the plasma velocities are assumed to vanish at
infinite distance to the bubble wall in the symmetric phase (positive
$z$), for which the default numerical value is set equal to
four times the bubble wall thickness $L_W$. The user can choose,
however, any other value. It was
checked numerically that the choice of $z_{max} = 4 L_W$ \cite{Basler:2019ici} for the boundary condition 
is on the one hand numerically stable and on the other hand 
does not significantly change the result. \s

As a second alternative variant of the transport equations, the plasma
velocities in \cref{Eq:TransportEquations} are expressed through the
derivatives of
\cref{Eq:TransportEquations:uh,Eq:TransportEquations:ut,Eq:TransportEquations:utc,Eq:TransportEquations:ub}. In
doing so, the derivatives of the thermal coefficients are dropped, moreover
derivatives of third or higher order are neglected, allowing to
express the plasma velocities as  
\begin{subequations}
  \begin{align}
    \partial_z u_{t,2} ={} & \frac{\partial_z S_t + 3 K_{4,t} \partial_z^2 \mu_{t,2}}{3\left(\Gamma_t^\mathrm{tot} + K_{6,t} \vw \partial_z m_t^2 \right)}  \,,\\
    \partial_z u_{t^c,2} ={} & \frac{\partial_z S_t + 3 K_{4,t} \partial_z^2 \mu_{t^c,2}}{3\left(\Gamma_t^\mathrm{tot} + K_{6,t} \vw \partial_z m_t^2 \right)}  \,,\\
    \partial_z u_{b,2} ={} & \frac{K_{4,b}}{\Gamma_b^\mathrm{tot}} \partial_z^2 \mu_{b,2}  \,,\\
    \partial_z u_{h,2} ={}& \frac{K_{4,h}}{\Gamma_h^\mathrm{tot}} \partial_z^2 \mu_{h,2} \,,
  \end{align}
  \label{Eq:EWBG:VelocityReplacement}
\end{subequations}
and the source term as 
\begin{align}
	\partial_z S_t ={} & \vw K_{9,t} \left( m_t^2 \partial_z^2 \theta  + \partial_z \theta \partial_z m_t^2  \right) \partial_z m_t^2 - 2 \vw K_{8,t} \partial_z m_t^2 \partial_z^2 \theta \,. 
\end{align}
These expressions for the plasma velocities allow to formulate the
transport equations as a differential system of equations of second
  order. As boundary conditions analogous conditions as in the
previous approach are chosen. The chemical potentials and their
derivatives are assumed to vanish at $z_{max}=4 L_W$. \s 

\Cref{eq:FH_MUBL} allows to determine the produced BAU
by evaluating the integral
\begin{align}
	\eta_\mathrm{B} =& \frac{405 \Gamma_{ws}}{4\pi^2 \vw g_{*} T} \int\limits_{0}^{\infty}  \mathrm{d}z \mu_{B_L}(z) \exp\left(- \frac{45\Gamma_{ws}}{4\vw} z\right) \label{Eq:Baryo}
\end{align}
numerically, where $g_{*}=106.75$ is the effective number of degrees of freedom in
the plasma \cite{Cline2000b,Cline2000a}, $\vw$ the wall velocity and
$\Gamma_{ws}$ is the weak sphaleron rate, given by 
\cite{Huet1995,Moore2000,Moore1997} 
\begin{align}
		\Gamma_{ws} ={}& 10^{-6} T \,.
\end{align}

\subsection{VEV-insertion Method/Fluid Equation \label{sec:via}}
As a second approach for the calculation of the BAU the VEV-insertion
method (\VIA) is chosen. The \VIA can be understood as an expansion in
$v(z)/T$ in which the fermionic two-point function is expanded. This
approach allows also to include further lighter quarks and leptons in
the transport equations. Leptons do not suffer from the
strong sphaleron suppression and their diffusion
  constant is significantly larger compared to that of the quarks,
  allowing for a more efficient diffusion process. Furthermore, in
  some specific models and parameter regions the lepton interaction
  rates are enhanced so that a sufficiently large BAU can be produced
  through leptons \cite{Chung:2009cb}.
In {\tt \ewBSMPT} the inclusion of the
dominant lepton contribution from the $\tau$ leptons and of the
contributions from the two
heaviest quarks, $t$ and $b$, is possible. \s

The implementation of the transport equations follows
Ref.~\cite{deVries:2017ncy,Vries2018} and is adapted accordingly. The chemical
potentials for the particles contributing to the
system of transport equations are
expressed as net chemical potentials, meaning that in the following
$\mu^{(x)}$ refers to the chemical potential of the
particles $x$ minus the one of their anti-particles. We denote the net
number densities\footnote{Note that the
    chemical potentials $\mu$ and 
  the particle densities $n$ are related through the high-temperature
  expansion, $n = \frac{T^2}{6}\kappa \mu +\mathcal{O}(\mu^3)$,
  with the statistical factor $\kappa$ defined below.} of the
left- and right-handed  quark/lepton and scalar degrees of
freedom by  
\begin{align}
    & n_q = n_{t_L} + n_{b_L}\,, && n_t = n_{t_R}\,, && n_b = n_{b_R}\,, \\\notag
    & n_l = n_{\nu_L} + n_{\tau_L}\,, && n_\tau = n_{\tau_R}\,, && n_\nu = n_{\nu_R}
\,,
\\ & \notag  n_{h_k} = n_{h^0_k}+n_{h^{\pm}_k}\,.
\end{align}
The index $k$ counts the doublets $\Phi_k=(h_k^\pm,h_k^0)$ considered
in the model under investigation. The strong sphaleron interactions
allow to relate the light quark densities via 
\beq
n_{q_1} = n_{q_2} = - 2 n_u = -2 n_d = -2n_s = -2n_c
\eeq
so that only one of them needs to be considered, which we choose to be $n_u$.
The transport equations including the dominant quark contributions of
$t,b$, the contribution of the $\tau$ lepton as well as the neutrino, 
the light quarks and the Higgs contributions are given by \cite{Vries2018}
\begin{subequations}
  \begin{align}
      & \partial_{\mu}j_q^{\mu} = + \Gamma_M^{(t)} \mu_M^{(t)} + \Gamma_M^{(b)} \mu_M^{(b)} + \Gamma_Y^{(t)} \mu_Y^{(t)}+ \Gamma_Y^{(b)} \mu_Y^{(b)} - 2 \Gamma_{ss} \mu_{ss} - S_{\cancel{CP}}^{(t)} - S_{\cancel{CP}}^{(b)}\\
      &\partial_{\mu}j_t^{\mu} = - \Gamma_M^{(t)}\mu_M^{(t)} - \Gamma_Y^{(t)}\mu_Y^{(t)} + \Gamma_{ss}\mu_{ss} + S_{\cancel{CP}}^{(t)}\\
      &\partial_{\mu}j_b^{\mu} = - \Gamma_M^{(b)}\mu_M^{(b)} - \Gamma_Y^{(b)}\mu_Y^{(b)} + \Gamma_{ss}\mu_{ss} + S_{\cancel{CP}}^{(b)}\\
      &\partial_{\mu}j_l^{\mu} = + \Gamma_M^{(\tau)}\mu_M^{(\tau)} + \Gamma_Y^{(\tau)}\mu_Y^{(\tau)} - S^{(\tau)}_{\cancel{CP}}\\
      &\partial_{\mu}j_\nu^{\mu} = 0\\
      &\partial_{\mu}j_\tau^{\mu} = -\Gamma_M^{(\tau)}\mu_M^{(\tau)} - \Gamma_Y^{(\tau)}\mu_Y^{(\tau)} + S_{\cancel{CP}}^{(\tau)}\\
      &\partial_{\mu}j_{h_k}^{\mu} = + \Gamma_Y^{(t)} \mu_Y^{(t)} -\Gamma_Y^{(b)} \mu_Y^{(b)}+ \Gamma_Y^{(u)}\mu_Y^{(u)}-\Gamma_Y^{(\tau)}\mu_Y^{(\tau)}\\
      &\partial_{\mu}j_u^{\mu} = + \Gamma_{ss} \mu_{ss}\,.
  \end{align}
  \label{VIA_Transport}
\end{subequations}
As noted above, here the distribution function of the up quark is chosen. \s 

The left-hand side of \cref{VIA_Transport} can be expressed in the
rest frame of the wall as 
\begin{equation}
  \partial_{\mu}j^{\mu}_i(x)\approx v_W \primed{n_i}- D_i \nabla^2 n_i \approx v_W\primed{n_i}-D_i n_i^{\prime\prime}\quad \cbrak{i=q,t,b,l,\nu,\tau,h_k,u}\,,
  \label{Fick_Law}
\end{equation}
with the diffusion constants for the quarks given by 
\beq
D_q = \frac{6}{T_c} \;.
\eeq 
The diffusion constants for the left-handed leptons are implemented as
\beq
D_l=\frac{100}{T_c}
\eeq 
and the diffusion constant for the right-handed
$\tau$ as 
\beq
D_{\tau} = \frac{380}{T_c} \;.
\eeq 
The single and double derivatives denoted by prime(s) are taken with
respect to the bubble wall distance $z$. 
The diffusion constant for the Higgs bosons is taken as 
\begin{equation}
  D_{h_k} = \frac{100}{T_c}\,.
\end{equation}
The strong sphaleron rate is given by 
\begin{equation}
  \Gamma_{ss}  = 14 \alpha_s^4 T_c\,,
\end{equation}
and the Yukawa rates of the quarks are implemented as \cite{Joyce:1994zn}
\begin{equation}
  \Gamma_Y^{(q)} = 0.19 \alpha_s y_q^2 T_c\,,
\end{equation}
where $\alpha_s$ denotes the strong coupling constant and $y_q$ the
zero-temperature Yukawa strength. The
Yukawa interaction rate for the $\tau$ lepton is given by  
\begin{equation}
  \Gamma^{(\tau)}_Y = 0.28 \alpha_w y_{\tau}^2 T_c\,,
\end{equation} 
with $\alpha_w=g_2^2/(4\pi)$, where
$g_2$ denotes the $\mathrm{SU}(2)$ gauge coupling, and the zero-temperature
Yukawa strength $y_\tau$. \s

The chemical potentials on the right-hand side of \cref{VIA_Transport}
are combinations of the net chemical potentials of the particles and
are given by  
\begin{align}
  &\mu_M^{(t)} =\cbrak{\frac{n_t}{\kappa_t} -
    \frac{n_q}{\kappa_{q}}}\,, && \mu_Y^{(t)} =
                                  \cbrak{\frac{n_t}{\kappa_t}-\frac{n_q}{\kappa_q}-\sum_k\frac{n_{h_k}}{\kappa_{h_k}} }\,,\\
  &\mu_M^{(b)}= \cbrak{\frac{n_b}{\kappa_b} - \frac{n_q}{\kappa_{q}}}\,, && \mu_Y^{(b)} = \cbrak{\frac{n_b}{\kappa_b}-\frac{n_q}{\kappa_q}+\sum_k\frac{n_{h_k}}{\kappa_{h_k}}}\,,\\
  &\mu_M^{(\tau)}= \cbrak{\frac{n_\tau}{\kappa_\tau} - \frac{n_l}{\kappa_{l}}}\,, && \mu_Y^{(\tau)} = \cbrak{\frac{n_\tau}{\kappa_\tau}-\frac{n_l}{\kappa_l}+\sum_k\frac{n_{h_k}}{\kappa_{h_k}}}\,,\\
  &\mu_{ss} = \cbrak{\frac{2 n_q}{\kappa_q} -\frac{n_t}{\kappa_t} - \frac{n_b}{\kappa_b}-\frac{8 n_u}{\kappa_{L}}-\frac{4 n_u}{\kappa_R}}\,,
\end{align}
where the chemical potentials are rescaled with a factor of
$6/T^2$. This factor is absorbed in the interactions rates. The
statistical factor $\kappa_L$ ($\kappa_R$) corresponds to the
statistical factor of a massless left (right)-handed quark. The
right-handed CP-violating source term is implemented as $(i=t,b,\tau)$
\begin{align}
  \label{eq:Scp}
  S_{\cancel{CP}}^{(i)} = \frac{N_c v_W}{\pi^2}\mathrm{Im}\cbrak{\primed{f}_if_i^*}\int dk\frac{k^4}{\omega^4} \left[-\frac{\Gamma_T}{2 \omega}+\frac{5 \Gamma_T}{4 \omega^2}\delta\omega + \cbrak{\frac{\Gamma_T}{\omega}-\frac{5\Gamma_T \delta\omega}{2 \omega^2} }n(\omega)\right.\\\nonumber
  +\left.\cbrak{\frac{-\omega^2}{2\Gamma_T}+\frac{\omega^4}{2k^2\Gamma_T}-\frac{\Gamma_T}{2}+\cbrak{\frac{\omega}{2\Gamma_T}+\frac{3\Gamma_T}{2\omega}}\delta\omega}\primed{n}(\omega)\right]\\\nonumber
  +\mathcal O\cbrak{\delta\omega^2;\cbrak{\frac{\Gamma_T}{T}}^2;n^{\prime\prime}}\,,
\end{align}
with the colour factor $N_c=3 (1)$ for top, bottom
  ($\tau$), and $k$ denoting the four-momentum of particle $i$. Note
that, for better readability, we neglect the index $i$ in
$n$, $k$ and $\omega$ (which is defined below). 
The Yukawa interaction strength $f_i$ corresponds to the interaction of
the left- and right-handed fermions in the Lagrangian 
\begin{equation}
  \mathcal L \supset - \frac{f_i\cbrak{T,\phi_b}}{\sqrt 2} \bar{\Psi}^i_L \Psi^i_R +\text{h.c.}\,,
  \label{param}
\end{equation}
depending on the background Higgs field configuration $\phi_b$
and thermal interactions. $f_i(T,\phi_b)$ can be split in one part
containing the contribution from the one-loop corrected mass at zero temperature
and another part containing the thermal mass contributions. We use the
following thermal masses for quarks ($q$) and leptons ($l$),
respectively, \cite{Postma:2019scv} 
\begin{subequations}  
\begin{align}
  &\cbrak{m_{T,R}^q}^2 = \cbrak{\frac{g_1^2}{18} + \frac{g_3^2}{6} + \frac{y_q^2}{8}}T^2\,,\\
  &\cbrak{\delta m^q}^2 \equiv \cbrak{m_{T,R}^q}^2-\cbrak{m_{T,L}^q}^2 =\cbrak{\frac{5 g_1^2}{96} - \frac{3g_2^2}{32} + \frac{y_q^2}{16}}T^2\,,\\
  &\cbrak{m_{T,R}^l}^2 = \cbrak{\frac{g_1^2}{8}  + \frac{y_l^2}{8}}T^2\,,\\
  & \cbrak{\delta m^l}^2 \equiv \cbrak{m_{T,R}^l}^2-\cbrak{m_{T,L}^l}^2 =\cbrak{\frac{3 g_1^2}{32} - \frac{3g_2^2}{32} + \frac{y_l^2}{16}}T^2\,,
\end{align}
\end{subequations}

with the SM gauge couplings $g_1,g_2,g_3$ and the zero-temperature
Yukawa couplings $y_q,y_l$. The gauge coupling $g_1$ is not GUT normalised.
We chose to parametrise the source terms and relaxation rates
$\Gamma_M^{(i)}$ in terms of
the respective right-handed mass and the mass difference $\delta m$ to
the corresponding left-handed mass. This allows us to expand the source
terms and relaxation rates in $\delta
m$. The dispersion of the left($L$)-/right($R$)-handed particle $i$
entering the source term is given by  
\begin{equation}
  \epsilon^i_{L/R} =\omega^i_{L/R} + \ii \Gamma_{T,L/R}^i \equiv
  \sqrt{(k^i)^2  +\cbrak{m^i_{T,L/R}}^2}-\ii \Gamma_{T,L/R}^i\,.
\end{equation}
The thermal width is approximately the same for all particles
$i=t,b,\tau$ and given by (the index $i$ can be dropped)
\begin{equation}
  \Gamma_{T,L}\approx\Gamma_{T,R}\approx\Gamma_{T}\approx 0.16 T\,.
\end{equation}  
Furthermore, the source term in \cref{eq:Scp} was expanded for small $\Gamma_T/T \ll 1$. The Fermi-Dirac distribution function is given by 
\begin{equation}
  n(x)=\cbrak{\mathrm{e}^x+1}^{-1}\,,
\end{equation}
and the expansion parameter $\delta\omega^i$ by 
\begin{equation}
  \delta\omega^i = \frac{\cbrak{\delta
      m^i}^2}{2\sqrt{(k^i)^2+\cbrak{m_R^{i}}^2}}\,.
\end{equation}
Analogously, the relaxation rates are expanded for small thermal widths
$\Gamma_T$ and small differences in the left- and right-handed thermal
masses and implemented as  
\begin{align}
  \Gamma^{(i)}_M = \frac{6}{T^2}\cdot \frac{N_c}{2\pi^2T}\vert f_i\vert^2 \int \frac{ dk k^2  }{\omega^2} \cbrak{-\frac{k^2}{\Gamma_T}+\frac{\omega^2}{\Gamma_T} + \frac{k^2\Gamma_T}{\omega^2} + \cbrak{\frac{k^2}{\omega\Gamma_T}-\frac{2 k^2 \Gamma_T}{\omega^3}}\delta\omega} h(\omega)\\\nonumber
  +\mathcal O\cbrak{\delta\omega^2;\cbrak{\frac{\Gamma_T}{T}}^2;h^{\prime}}\,,
\end{align}
with 
\beq
h(x) =
\mathrm{e}^x/\cbrak{\mathrm{e}^x+1}^2 \;,
\eeq 
and where again we suppressed the index $i$ in $k$, $\omega$ and
$\delta \omega$ for better 
readability. The scale factor of $\frac{6}{T^2}$ from the high-temperature
expansion of the chemical potentials is absorbed in the relaxation
rates. The high-temperature expansion of the chemical potential $\mu$ yields
the following relation between the chemical potential $\mu$ of the particle
and its number density $n$, 
\begin{equation}
  n = \frac{T^2}{6}\mu \kappa + \mathcal O(\mu^3)\,,
  \label{high_temp_chemical_potential}
\end{equation}
with the statistical factor for fermions ($F$,+) and bosons ($B$,$-$), 
respectively, 
\begin{equation}
  \kappa(x) = \kappa(0) \frac{c_{F,B}}{\pi^2} \int\limits_{m/T}^{\infty} dx \frac{x \mathrm{e} ^x}{\cbrak{\mathrm{e}^x\pm 1}^2} \sqrt{x^2-m^/T^2}\,,
\end{equation}
where the numerical values for the normalisation are
  given by $c_F = 6$ and $c_B=3$. \s

All integrals over the momentum $k$ above are evaluated numerically in
{\tt \ewBSMPT} in the range of $k\in\left[0,10^4\right]$. The upper
boundary is chosen such that a numerically stable result can be
obtained and at the same time the result does not depend significantly
on the chosen cut. The differential system of equations in
\cref{VIA_Transport} is solved with the help of the {\tt boost}
library at a given distance $z$. Analogously to the \FH approach the
boundary condition is chosen such 
that the chemical potentials vanish at $z_{max}=4L_W$. The resulting
grid of the sum of all left-handed quarks and leptons \cite{Vries2018},
\begin{equation}
  n_L^0(z) \approx  \cbrak{n_q-4 n_u + n_l}\,,
\end{equation}
is interpolated as cubic spline and finally, the following differential equation
is integrated to determine the produced baryon density $n_B$
\begin{equation}
  v_W n_B^{\prime} - D_q n_B^{\prime\prime} = -
  \Gamma_{ws}\cbrak{\frac{3}{2}n_L^0 + \mathcal R n_B}\,.
  \label{ODE::BARYO}
\end{equation}
The electroweak sphaleron transition rate is implemented as 
\begin{equation}
  \Gamma_{ws} = 6 \kappa \alpha_w^5 T_c\approx 5 \cdot 10^{-6} T_c\,,
\end{equation}
with the numerical factors $\kappa\approx 20 $ and 
$\mathcal{R}=\frac{15}{4}$.

\section{Program Description \label{sec:program}}
In this section, we describe the system requirements for {\tt
  \ewBSMPT} and provide a guide for installing and using the program. 

\subsection{System Requirements}
The program was developed and tested on OpenSuse 42.2, Ubuntu 14.04,
Ubuntu 16.04, Ubuntu 19.10, Ubuntu 20.04 and Mac 10.13 
with {\tt g++ v6.2.1}, {\tt g++ v.7.2.1}, {\tt g++ v.9.3.0} and {\tt g++ 10.2.0}. The program can be downloaded at:
\begin{center}
  \download
\end{center}
For a working version of {\tt \ewBSMPT} the following dependencies
are required:
\begin{itemize}
  \item {\tt CMake v3.13} or higher. It can be installed either
    through {\tt pip} or directly from
    \url{https://cmake.org/download/}.
  \item The GNU Scientific Library ({\tt GSL}) \cite{GSL} v2.1 or higher is
        assumed to be installed in PATH. {\tt GSL} is
        required for the calculation of the Riemann-$\zeta$ functions, the
        double factorial and for the minimization. 
  \item {\tt Eigen3 v3.3} \cite{eigenweb} or
    higher is used to calculate the eigenvalues and eigenvectors of the mass matrices.
  \item {\tt Boost 1.66} \cite{Boost1.66} is required for the
    computation of the BAU. For the calculation of the phase
    transition the {\tt Boost} library is optional.  
\end{itemize}
Furthermore, at least one of the following two libraries has to be provided
\begin{itemize}
  \item {\tt libcmaes} \cite{CMAES}. Per default {\tt BSMPT} will
    install {\tt libcmaes} in the build directory, which is
    recommended for proper links to {\tt Eigen3}. 
  \item {\tt NLopt} \cite{NLopt} using the {\tt DIRECT\_L}
    \cite{NLOPTDirectL} algorithm. As an additional option for the
    minimization we provide a link to {\tt NLopt}.
\end{itemize}

\subsection{Download}
The latest stable version of the program package {\tt \ewBSMPT} can be
obtained from
\begin{center}
  \download\ .
\end{center} 
We also provide
  a Doxygen documentation there which incorporates all functions and parameters
  of \texttt{BSMPT v2}. The user can either clone the repository or download the
package as a zip archive. The chosen directory of the user is referred
to as {\tt \$BSMPT} in the following. The main directory contains the
following subdirectories:\s 

%
\begin{longtable}{ll}
  {\tt example} & In this subfolder sample input files needed for the binaries are provided. \\[0.1cm]
  {\tt manual}  & This subfolder contains a copy of the manual of {\tt BSMPT} 
  kept up to date                                                                                     \\ & with changes in the code.
  Additionally, we include the {\tt changelog}                                                        \\ & file
  documenting corrected bugs and modifications of the
  program.                                                                                            \\[0.1cm]
  {\tt sh}      & In this subdirectory the python files
  {\tt prepareData\_XXX.py} (XXX= R2HDM,                                                                    \\
                & C2HDM, N2HDM) are provided allowing to order the
  data sample according to                                                                            \\
                & the input requirements.                                                             \\[0.1cm]
  {\tt src}     & This subfolder contains the source files of the code.                               \\[0.1cm]
  {\tt include} & This subfolder contains the header files of the code.                               \\[0.1cm]
  {\tt tools}   & This folder contains configurations for the
                  installation with {\tt cmake}.                \\[0.1cm]
  {\tt maple}   & This folder contains a Maple worksheet template which can be used to calculate \\ & the  necessary tensors to implement a new model in {\tt BSMPT}.\\[0.1cm]
\end{longtable}

\vspace*{0.2cm}
\noindent
The subfolders of {\tt src} contain the following files:
\vspace*{0.2cm}

\begin{tabular}{ll}
  \hspace*{-0.32cm} {\tt src/minimizer}           & This directory contains the minimization routines.             \\
  \hspace*{-0.32cm} {\tt src/models}              & This directory contains all implemented
  models.                                                                                                                                                                \\ &  If a new model is added it must
  be added in this folder.                                                                                                                                              \\ & A template class 
   with instructions on how to add a new\\ & 
  model and the file {\tt SMparam.h} with the Standard Model 
  \\ & parameters are provided.                                                                                                                     \\
  \hspace*{-0.32cm} {\tt src/prog}                & This directory contains the source code for
  the executables.                                                                                                                                                       \\
  \hspace*{-0.32cm} {\tt src/ThermalFunctions}    & This directory contains the implementation of the thermal integrals.                                                 \\
  \hspace*{-0.32cm} {\tt src/WallThickness}       & This directory
                                                    contains the
                                                    calculation of the
                                                    wall thickness.                                                            \\
  \hspace*{-0.32cm} {\tt src/Kfactors}            & This directory contains the calculation and interpolation of the \\ &thermal transport coefficients $K_i$ used in one of the
  \\&implemented calculations of the BAU.        \\
  \hspace*{-0.32cm} {\tt src/baryo\_calculations} & This directory
                                                    contains the
                                                    implementation of the different \\ & approaches to calculate the BAU. \\
\end{tabular}

\subsection{Installation\label{sec:Installation}}
The compilation requires a {\tt C++} compiler supporting the {\tt C++14} standard.
To install {\tt \ewBSMPT} the following steps have to be performed:
\begin{enumerate}
  \item In the directory {\tt \home} call 
\begin{lstlisting}{style=simplestyle}
  mkdir build && cd build 
\end{lstlisting}
  \item In order to generate the makefile call 
\begin{lstlisting}{style=simplestyle}
  cmake [OPTIONS] ..
\end{lstlisting}
in {\tt\$BSMTP/build} with the following possible options
          \begin{itemize}
            \item {\tt CXX=C++Compiler} If not given the default compiler will be used.
            \item {\tt
                -DEigen3\_DIR=/path/to/eigen3Installation/share/eigen3/cmake}
Sets the path to the {\tt Eigen3} installation. Can be used in case
{\tt Eigen3} is not found automatically.
            \item {\tt -Dlibcmaes\_ROOT=/path/to/libcmaesInstallation} if libcmaes is installed previously. 
            \item {\tt -DUseLibCMAES=ON/OFF}, where the default is ON. Switch to decide if libcmaes should be used or not.
            \item {\tt -DNLopt\_Dir=/path/to/NLoptInstallation} This gives the corresponding path to the NLopt installation if NLopt is not found automatically.
            \item {\tt -DUseNLopt=ON/OFF}, default is ON. Switch to decide if NLopt should be used or not.
            \item {\tt -DBoost\_ROOT=/path/to/boost} This is only
              necessary if Boost is not found automatically through
              {\tt cmake}.
          \end{itemize}
A complete call would look like: 
\begin{lstlisting}{style=simplestyle}
  CXX=C++Compiler cmake -DEigen3_DIR=/path/to/eigen3Installation/share/eigen3/cmake -Dlibcmaes_ROOT=/path/to/libcmaesInstallation -DNLopt_Dir=/path/to/NLoptInstallation -DBoost_ROOT=/path/to/boost ..
\end{lstlisting}
  \item Subsequently call \\[0.1cm]
\begin{lstlisting}[language=bash]
  cmake --build .
\end{lstlisting}
If the installation is successful, the executables {\tt BSMPT}, {\tt
  CalcCT}, {\tt NLOVEV}, {\tt VEVEVO}, {\tt Test}, {\tt TripleHiggsCouplingsNLO}, {\tt
  CalculateEWBG}, {\tt PlotEWBG\_vw}, and {\tt PlotEWBG\_nL} are generated in \url{$BSMPT/build/bin/}.
\item (Optional) Furthermore, we provide a detailed documentation of each function implemented in {\tt \ewBSMPT} through doxygen. To generate the doxygen documentation call in {\tt \$BSMPT/build} 
\begin{lstlisting}
  cmake --build .  -t doc
\end{lstlisting} 
Afterwards, the documentation can be found as \url{$BSMPT/build/html/index.html}. Note that for the generation of the documentation the program package doxygen is required.
\item (Optional) We provide a unit-test, allowing the users to test their current installation of {\tt\ewBSMPT}. To perform the unit-test call in \url{$BSMPT/build/}
\begin{lstlisting}
  ctest 
\end{lstlisting}
The test calls subsequently test frames checking for
  various quantities \textit{e.g.} the NLO VEV configuration of a
  known benchmark scenario. So far, there are test frames available
  for the C2HDM, N2HDM and R2HDM. A relative agreement of the order
  $\mathcal{O}\cbrak{10^{-4}}$ is demanded in each tested
  quantity. Note that the effective potential is not sensitive to the
  sign of the VEV configuration due to the $SU(2)$ symmetry,
  hence the output of the minimization routines might differ in the
  sign of the VEVs. The absolute values, however,
  must be in agreement with the default values
  provided for the test frames. \s

\end{enumerate}

\subsection{Program Usage}
In the following the different executables of {\tt \ewBSMPT} are described. All executables of {\tt \ewBSMPT} need the following additional input parameters
\begin{itemize}
  \item {\it Model}: Input parameter to choose the model. Implemented are the CP-violating 2HDM ("c2hdm"), CP-conserving 2HDM ("r2hdm") and the CP-conserving N2HDM ("n2hdm"). With the release of {\tt \ewBSMPT} also the complex singlet extension ("cxsm") is added. 
  \item {\it Inputfile}: sets the path and the name of the input file.
        The program expects the first line in the input file 
        to be a header with the column names. Every
        following line then corresponds to the input of one particular
        parameter point. The parameters are required to be those of
        the Lagrangian in the interaction basis. If a different format
        for the input parameters is desired one needs to adapt the function
          {\tt ReadAndSet} in the corresponding model file in
          {\tt \home/src/models}. Note, that the program
        expects the input parameters to be separated by a tabulator. In the
        folder {\tt \$BSMPT/sh/}, we provide {\tt python} scripts that prepare
        the data accordingly. 
  \item {\it Outputfile}: sets the path and the name of
        the generated output file. We note, that the
        program does not create new folders so the user has to ensure the existence of the output directory. 
\end{itemize}
Furthermore, every executable has a help menu which can be called by 
\begin{lstlisting}{style=simplestyle}
  ./executable --help
\end{lstlisting}
In addition to the given parameters here, the help menu will show an
additional method for calling the executables with more options,
e.g. turning off the minimizers at runtime. 
It is possible to switch the treatment of the thermal corrections
from the default choice, referred to as Arnold Espinosa method, to
the alternative method, referred to as Parwani method, by changing the
variable {\tt C\_UseParwani} to true in the file  
\begin{center} 
{\tt \home/include/BSMPT/models/ClassPotentialOrigin.h}.
\end{center}
The changes have to be applied to the binaries through recompiling the code with {\tt make} in the directory {\tt \$BSMPT/build}.
In the following subsections we list the binaries of {\tt
  \ewBSMPT}. \s

\subsubsection{Test}
{\tt Test} provides a test binary to automatically check for
implementation errors \textit{e.g.} for new model implementations or
changes in the source code.  
The following checks are performed by the binary:
\begin{itemize}
  \item It checks if the tree-level minimum is given by the input
    value of $v\approx 246.22$~GeV within the numerical precision. 
  \item It compares between the masses of the quarks and gauge bosons given in the {\tt SMparam.h} file and the internally calculated numerical values.
  \item It minimizes the tree-level potential and checks if the input
    minimum is the global minimum. 
  \item It checks if the implemented simplified tree-level and
    counterterm potentials yield the same results as the generic potential.
  \item It checks if the number of labels in {\tt addLegendCT}, {\tt
      addLegendVEV}, {\tt addLegendTemp} matches with the required
    number of counterterm parameters and VEVs.
  \item It checks if the higher-order Higgs boson
    masses obtained with the implemented counter\-term potential are the
    same as the tree-level ones. 
\end{itemize}
The test binary can be executed with\\[0.1cm]
\begin{kasten*}{Test}
  ./bin/Test Model Inputfile Line
\end{kasten*} \\
{\it Model} specifies the model under investigation. The line number of
the parameter point which should be used for the test is given by {\it
  Line}. The results and status of the test will be prompted as
terminal output. An example call for the C2HDM would look like
\begin{lstlisting}
  ./bin/Test c2hdm ../example/C2HDM_Input.dat output.csv 2 
\end{lstlisting}
where the second line of \texttt{C2HDM\_Input.dat} would be used. 
\s

\noindent In the following, we will list the executables and their individual
function. Some of these executables are already present in the previous
version of {\tt BSMPT}. For completeness, we list them here as
well. The new executables in {\tt BSMPT} are {\tt CalculateEWBG}, {\tt
  PlotEWBG\_vw}, and {\tt PlotEWBG\_nL} 
described in
\cref{sec::binCalculateEWBG,sec::binPlotEWBGvw,sec::binPlotEWBGnL}
in the following.

\subsubsection{{\tt BSMPT}}
\label{sec::binBSMPT}
{\tt BSMPT} is the executable calculating the strength of the
electroweak phase transition of a given parameter point
in {\it Inputfile}. The binary is called by\\[0.1cm]
\begin{kasten*}{BSMPT}
./bin/BSMPT Model Inputfile Outputfile LineStart LineEnd
\end{kasten*} \\
Here and wherever they appear in the command lines in the following,
{\it LineStart} and {\it LineEnd} set the line numbers of the first and the last parameter
  point provided in {\it Inputfile} for which the various
  quantities are to be calculated by {\tt \ewBSMPT}.
Note that the first line of the input file is expected to be
the header, so that the first parameter point should have the ${\it
  LineStart}=2$. To calculate a single parameter point {\it LineStart}
and {\it LineEnd} have to be equal. The results of {\tt BSMPT} are
written in {\it Outputfile} where the parameter point read in is
extended by the results of the calculation. Also the legend from the
input file is extended by the additional legend needed for the
output. Every parameter point in the range of {\it LineStart} to {\it
  LineEnd} will be calculated and written to {\it Outputfile}
regardless if the parameter point provides an SFOEWPT or not. 
An example call for the C2HDM would look like
\begin{lstlisting}
./bin/BSMPT c2hdm ../example/C2HDM_Input.dat output.csv 2 2 
\end{lstlisting}
where the first data line of
  \texttt{C2HDM\_Input.dat} would be used. The output file
  \texttt{output.csv} should be the same as
  \texttt{example/C2HDM\_Input.dat\_BSMPT} up to numerical
  fluctuations.\s 

The column {\tt omega\_c/T\_c} provides the information if the
respective parameter points fulfill all checked conditions. If one
parameter point fails a check the column  is set to the following
values for this point: 
\begin{itemize}
  \item   {\tt omega\_c/T\_c}= -1: The VEV configuration is not
    vanishing at $T=300\gev$ and therefore no SFOEWPT (in a single
    step) is possible. 
  \item  {\tt omega\_c/T\_c}= -2: The next-to-leading order (NLO) VEV
    at zero temperature is 
    not equal to the tree-level VEV, hence the parameter is not NLO
    stable and neglected.
  \item {\tt omega\_c/T\_c}=-3: The parameter point provides
    electroweak VEV configurations above
    $255\gev$ and is therefore
    neglected due to unphysical behaviour.\footnote{We chose the value of 255~GeV to account
        for potential fluctuations in the numerical result.}  
  \item  {\tt omega\_c/T\_c}=-4: If the parameter point provides
    $\omega_c/T_c < \text{{\tt C\_PT}}$ the output is set to -4. Note
    that the default value of {\tt C\_PT} is set to zero. In order to
    obtain in the output file only parameter points that provide an
    SFOEWPT,  {\tt C\_PT} has to be set to 1.  
  \item {\tt omega\_c/T\_c}=-5: The parameter point provides a
    vanishing or divergent VEV configuration at zero temperature and
    therefore the calculation is aborted.  
\end{itemize}
If $\text{{\tt omega\_c/T\_c}}>0$, the parameter point has passed all
tests. Note that for parameter points with an SFOEWPT the value has to be
greater than one. Note, that the reason for errors can be multiple and
be due to implementation errors or have a physics reason or both so
that only the error itself will be reported. 

\subsubsection{CalcCT}
\label{sec::binCalcCT}
{\tt CalcCT} is the executable for the calculation of the
counterterms for a given parameter point. It is executed through the
command line \\[0.1cm]
\begin{kasten*}{CalcCT}
  ./bin/CalcCT Model Inputfile Outputfile LineStart LineEnd
\end{kasten*}\\
For each line,
{\it i.e.}~each parameter point, the various counterterms of the model
are calculated. They are written out in the output file which contains
a copy of the parameter point and appended to it in the same line the
results for the counterterms. The first line of the output file
contains the legend describing the entries of the various columns.
An example call for the C2HDM would look like
\begin{lstlisting}
  ./bin/CalcCT c2hdm ../example/C2HDM_Input.dat output.csv 2 2 
\end{lstlisting}
The output file \texttt{output.csv} should be the same as
\texttt{example/C2HDM\_Input.dat\_CalcCT} up to numerical
fluctuations.

\subsubsection{NLOVEV}
\label{sec::binNLOVEV}
{\tt NLOVEV} is the executable calculating the
global minimum of the loop-corrected effective potential at $T=0$~GeV
for every point between the lines {\it LineStart} and {\it LineEnd} to
be specified in the command line for the execution of the program: \\[0.1cm]
\begin{kasten*}{NLOVEV}
  ./bin/NLOVEV Model Inputfile Outputfile LineStart LineEnd
\end{kasten*}\\
The output file contains the information
on the parameter point for which the computed values at zero
temperature of the NLO VEVs (in GeV) are appended in the same line, namely $v(T=0)$
and the individual VEVs $\omega_k (T=0) \equiv v_k$
($k=1,...,n_v$), where $n_v$ is the number the VEVs
  present in the investigated model. The first line of the output file again details the
entries of the various columns. Note, that it can happen that
the global minimum $v(T=0)$, that is obtained from the NLO effective potential,
is not equal to $v=246.22$~GeV any more. By writing out also
$v(T=0)$ the user can check for this phenomenological constraint.
An example call for the C2HDM would look like
\begin{lstlisting}
  ./bin/NLOVEV c2hdm ../example/C2HDM_Input.dat output.csv 2 2 
\end{lstlisting}
The output file \texttt{output.csv} should be the same as
\texttt{example/C2HDM\_Input.dat\_NLOVEV} up to numerical
fluctuations. 
\subsubsection{VEVEVO}
\label{sec::binVEVEVO}
{\tt VEVEVO} provides the executable calculating the global minimum of
the loop-corrected effective potential between the temperatures {\tt
  TemperatureStart} and {\tt TemperatureEnd}. The binary is called
with\\[0.1cm] 
\begin{kasten*}{VEVEVO}
  ./bin/VEVEVO Model Inputfile Outputfile Line TemperatureStart TemperatureStep TemperatureEnd
\end{kasten*}\\
The output file contains the used parameter point extended with the
values of the NLO VEVs at the given temperature. For each temperature
step with a stepsize {\tt TemperatureStep} a line is added to the
output file. 
An example call for the C2HDM would look like
\begin{lstlisting}
  ./bin/VEVEVO c2hdm ../example/C2HDM_Input.dat output.csv 2 0 10 160
\end{lstlisting}
where the first data line of
  \texttt{C2HDM\_Input.dat} would be used and the temperature would be
  varied from $0~\mathrm{GeV}$ to $160~\mathrm{GeV}$ with a stepsize
  of $10~\mathrm{GeV}$.  
  The output file \texttt{output.csv} should be the same as
  \texttt{example/C2HDM\_Input.dat\_VEVEVO} up to numerical
  fluctuations. 

\subsubsection{TripleHiggsCouplingsNLO}
\label{sec::binTripple}
{\tt TripleHiggsCouplingsNLO} is the executable of the program that
calculates the triple Higgs couplings, derived from the third
derivative of the potential with respect to the Higgs fields, for every point between
the lines {\it LineStart} and {\it LineEnd} to be specified in the
command line: \\[0.1cm]
\begin{kasten*}{TripleHiggsCouplingsNLO}
  ./bin/TripleHiggsNLO Model Inputfile Outputfile LineStart LineEnd
\end{kasten*}\\
The output file contains the trilinear
Higgs self-couplings derived from the tree-level potential, the
counterterm potential and the Coleman-Weinberg potential at $T=0$ for
all possible Higgs field combinations. The total NLO trilinear Higgs self-couplings are
then given by the sum of these three contributions. The first line
of the output file describes the entries of the various columns. 
An example call for the C2HDM would look like
\begin{lstlisting}
  ./bin/TripleHiggsNLO c2hdm ../example/C2HDM_Input.dat output.csv 2 2 
\end{lstlisting}
The output file \texttt{output.csv} should be the same as
\texttt{example/C2HDM\_Input.dat\_TripleHiggsNLO} up to numerical
fluctuations. 

\subsubsection{CalculateEWBG \label{sec:calcewbg}}
\label{sec::binCalculateEWBG}
{\tt CalculateEWBG} is the executable of the program that calculates
the BAU for each parameter point between {\it LineStart} and {\it
  LineEnd} in the different approaches. The command line
reads:\\[0.1cm] 
\begin{kasten*}{CalculateEWBG}
  ./bin/CalculateEWBG Model Inputfile Outputfile LineStart LineEnd EWBGConfigFile
\end{kasten*}\\
So far, the routines for the calculation of the BAU are only implemented
for the C2HDM. This means that at present {\it Model} can only be set
to C2HDM. {\tt EWGBConfigFile} allows to determine which approach
is used for the calculation of the BAU. A sample config file is
given in {\tt \$BSMPT/example/EWBG\_config.txt} and can be modified
accordingly. 
The default config file has the following form
\begin{lstlisting}
    ### config file for the EWBG calculation (yes/no)

    VIA Ansatz only including the top quark in the transport equations
    Include:    yes
    
    VIA Ansatz including the top and bottom quark in the transport equations
    Include:    yes
    
    VIA Ansatz including the top and bottom quark and the tau lepton in the transport equations
    Include:    yes

    Via Ansatz treating the bottom quark massive
    Massive:    yes
    
    FH Ansatz with the plasma velocities
    Include:    yes
    
    FH Ansatz with the plasma velocities replaced through the second derivatives
    Include:    no
    
  \end{lstlisting} 
By changing {\tt yes/no} it is possible to switch on/off the different approaches in the calculation of the BAU. The spacing in front of yes/no has to be adhered to. The following approaches can be chosen: 
\begin{itemize}
  \item \VIA approach with only the top quark included in the transport equations,
  \item \VIA approach with top and bottom quarks included in the transport equations,
  \item \VIA approach with top and bottom quarks and $\tau$ lepton in
    the transport equations, 
 \item \VIA approach with the bottom quark treated massive (yes) or
   massless (no), 
  \item \FH approach with plasma velocities included,
  \item alternative \FH approach with plasma velocities replaced using \cref{Eq:EWBG:VelocityReplacement}.
\end{itemize}
An example call for the C2HDM would look like
\begin{lstlisting}
  ./bin/CalculateEWBG c2hdm ../C2HDM_Input.dat output.csv 2 2 ../example/EWBG_config.txt
\end{lstlisting}
The output file \texttt{output.csv} should be the same as \texttt{example/C2HDM\_Input.dat\_EWBG} up to numerical fluctuations.
\subsubsection{PlotEWBG\_vw}
\label{sec::binPlotEWBGvw}
The binary {\tt PlotEWBG\_vw} allows to calculate the BAU as a function of the bubble wall velocity $v_W$. Note that the wall velocity is treated as an additional input parameter in the implemented approaches. The binary is called by \\[0.1cm]
\begin{kasten*}{PlotEWBG\_vw}
  ./bin/PlotEWBG\_vw Model Inputfile Outputfile Line vwMin vwStepsize vwMax EWBGConfigFile
\end{kasten*}\\
So far only the C2HDM is implemented so that {\it Model} has to be
chosen accordingly.
The line number {\it
  Line} determines the parameter point in the given input file. The BAU
will be calculated for this parameter point for equidistant wall velocities
between {\it vwMin} and {\it vwMax} with a stepsize of
{\it vwStepsize}. {\tt EWBGConfigFile} allows to specify which
approach is used in the calculation. A sample config
file can be found in \$BSMPT/example/EWBG\_config.txt. The
options are the same as in the binary {\tt CalculateEWBG}, cf. \cref{sec::binCalculateEWBG}.
An example call for the C2HDM would look like
\begin{lstlisting}
  ./bin/PlotEWBG_vw c2hdm ../example/C2HDM_Input.dat output.csv 2 0.05 0.01 0.15 ../example/EWBG_config.txt
\end{lstlisting}
The output file \texttt{output.csv} should be the same as \texttt{example/C2HDM\_Input.dat\_PlotEWBG\_vw} up to numerical fluctuations.

\subsubsection{PlotEWBG\_nL}
\label{sec::binPlotEWBGnL}
The binary {\tt PlotEWBG\_nL} determines the left-handed fermion
densities in front of the bubble wall as a function of the
wall distance $z$. The binary is called by \\[0.1cm]
\begin{kasten*}{PlotEWBG\_nL}
  ./bin/PlotEWBG\_nL Model Inputfile Outputfile Line vw EWBGConfigFile
\end{kasten*}\\
We remind that at present {\it Model} can only be set to the C2HDM.
The line number of the parameter point for
which the calculation is performed is set by {\it Line}, and the input
wall velocity $v_W$ by {\it vw}. In 
{\tt EWBGConfigFile} the user can choose which approach for the
calculation of the BAU should be 
used, as explained in \cref{sec::binCalculateEWBG}. A sample input file is provided in  
  \$BSMPT/example/EWBG\_config.txt.  In the output file for
each step of the distance $z$ an additional line is added with the
corresponding results for the left-handed densities of the chosen
approaches.  
An example call for the C2HDM would look like
\begin{lstlisting}
  ./bin/PlotEWBG_nL c2hdm ../example/C2HDM_Input.dat output.csv 2 0.1 ../example/EWBG_config.txt
\end{lstlisting}
The output file \texttt{output.csv} should be the
  same as \texttt{example/C2HDM\_Input.dat\_PlotEWBG\_nL} up to
  numerical fluctuations.

\section{Discussion of Approaches, Options and
    Approximations \label{sec:discussion}}

We have implemented in {\tt BSMPT v2} two different approaches for the formulation
  of the quantum transport equations, the {\tt FH} approach and the
  {\tt VIA } method, for which we also allow different options, {\it
    cf.}~Sec.~\ref{sec:calcewbg}. We furthermore use various
  approximations. We plan to investigate in 
  detail in a forthcoming publication the impact of the different approaches and
  approximations on the BAU. We discuss them more generally here in
  order to give some guidance to the user of the code. \s

\noindent
{\it The two different approaches for the computation of the BAU} \\
In the literature both the {\tt FH} and the {\tt VIA} method have been
used to compute the BAU. Thus the BAU for the Minimal Supersymmetric Extension of
the Standard Model (MSSM) has been obtained with the {\tt VIA} method
in \cite{Carena:1997gx} {\it e.g.}, and with the {\tt FH} approach in
\cite{Cline:1997vk,Cline:2000nw,Cline:2000kb}. While in both
approaches the particle densities that contribute to the BAU are
obtained from the Boltzmann equations, there is no agreement on their
precise form. A systematic comparison of the two
methods for the C2HDM is still
missing. Ref.~\cite{Konstandin:2013caa}, where the quantum transport
equations were derived from first principles in the Schwinger-Keldysh 
formalism, presents a brief general comparison with the {\tt FH} and {\tt VIA}
approach mentioning also shortcomings inherent in each approach. Furthermore, a quantitative comparison between the different approaches is given for the MSSM. The author 
also points out that in the {\tt VIA} method it is straightforward to
include several flavours. In 
\cite{Cline:2020jre}, a comparison was performed for a
prototypical model of CP violation in the wall. In agreement with the
results obtained for the MSSM, the authors found that the {\tt VIA} estimate
of the BAU is significantly larger. The authors of
\cite{Cline:2020jre}, who apply the {\tt FH} approach to investigate
the BAU, argue that these semiclassical results are more
reliable. In view of missing further investigations we regard the
question on which approach to be used as still an open issue and
therefore offer both options in our code. We point out to the user,
however, that in our experience the results obtained for the BAU in the {\tt VIA}
approach are considerably larger than those of the {\tt FH} method, so
that the computed BAU result should be rather regarded as an estimate and
taken as an indication if the model under investigation might lead to
sufficient BAU in principle. \s

\noindent
{\it Options provided for the two different approaches} \\
In the configuration file {\tt EWGB\_config.txt} presented in
Sec.~\ref{sec:calcewbg} different options can be chosen to compute the
BAU. As stated in \cite{Konstandin:2013caa} the {\tt VIA} approach can
easily be extended to additional flavours. The first three options
given in the configuration file allow in the {\tt VIA} approach, in the order in which they
appear, the inclusion of the top quark only,
the additional inclusion of the bottom quark, and the inclusion of the
top, bottom and tau. By choosing and comparing the different options
the user can study the impact of additional flavours and get a feeling
on possible directions for model building. The fourth option relates
to the {\tt VIA} approach in general. In case the bottom quark is
included, its mass can be set to zero by setting the corresponding flag to 'no'. The
last two options relate to two possible variants of the transport
equations in the {\tt FH} approach. In the first variant
the plasma velocities are used, in the second they are expressed
through derivatives, {\it
  cf.}~Eq.~(\ref{Eq:EWBG:VelocityReplacement}). For a comparison of the two variants,
see~\cite{phdbasler}. Offering the choice among these options allows the user to
study the impact of the two approaches and, in the {\tt VIA} approach, of
the different fermion configurations. For the reasons stated above, we do not give
a recommendation for which option to be used. \s

\noindent
{\it Small Wall Velocities} \\
As stated in Sec.~\ref{sec:fh} we apply the approximation of small
wall velocities in the {\tt FH} approach. Recently,
\cite{Cline:2020jre} rederived the fluid equations without making this
approximation. It was found that the BAU suppression is a smooth
function of the wall velocity and the sound speed barrier can safely
be crossed. The authors furthermore pointed out some mistakes in the previous
derivation of the {\tt FH} approach. They presented also a numerical
comparison of the BAU with their new results and their previous {\tt FH}
implementation. It is found that both results agree for small wall
velocities. With increasing $v_W$ they start to differ reaching a
deviation below 30\% for $v_W=0.1$ in the predicted BAU for the prototypical model
studied in the paper. While the difference depends on the specific
model, as a rough guideline we recommend not to use wall velocities
above 0.1 when applying our code for the {\tt FH} method. We will
implement the new results of 
\cite{Cline:2020jre} which requires some major changes in the code in
the next upgrades so that also larger wall velocities can be
applied. In the meantime, the user can also study the effect of going
slightly beyond 0.1. As long as the results for the BAU do not change
drastically, the obained numbers should still provide a reasonable
estimate of the BAU. The {\tt VIA} method does not utilize any explicit expansion
in small  $v_W$, it still assumes, however, small bubble wall dynamics
in the derivation of the quantum transport equations. In this sense,
we also advise the user to use bubble wall velocities of about the
same magnitude as in the {\tt FH} method. \s

\noindent
{\it Impact of the wall profile} \\
As discussed in Sec.~\ref{sec:wallprofile} we approximate the bubble
wall profile by the kink profile, an approach that is frequently used
in the literature. We also described in Sec.~\ref{sec:wallprofile} how
we obtain the barrier height. We furthermore varied the
  wall thickness for specific benchmark points and found that the impact on the BAU
  was in the percentage region. In order to estimate the error made
  by our approximation of the wall profile we would need to compare it
  with the bounce solution. \s

\noindent
{\it Usage of the critical temperature} \\
In {\tt BSMPT v2} we are not able to estimate the nucleation
temperature $T_N$  and the bounce solution. Instead we
determine the wall thickness and bubble profile at the critical
temperature. One could argue to provide the possiblity to use the
nucleation temperature obtained by a different code. We deliberately
refrain, however, from including the nucleation temperature from a
different code. The reason is that this would be 
inconsistent. 
Changing the temperature from the critical to the
nucleation temperature would yield a different vacuum structure of the
potential. The new vacuum structure would prevent the usage of the
{\tt BSMPT} algorithms, since these require the degenerate vacua at
the critical transition point.
Remark that we define and apply the VEV(s) determined
at the critical temperature. In other words, the results obtained with
$T_N$ instead of $T_c$ would imply inconsistent solutions. We
therefore do not provide the option to use $T_N$ obtained from a
different code. The consistent solution requires the future determination of
$T_N$ and the bounce solution within {\tt BSMPT}. An approximate
feeling of the influence of our approximation can be obtained by
investigating the resulting BAU for given $\xi_c$ values in the model
under investigation (at present only the C2HDM is possible) in a large
number of allowed parameter points of the model. This requires a large
scan and a detailed phenomenological investigation that we plan to
perform in a forthcoming paper.

\section{How to add a New Model \label{sec:newmodels}}
In the following section a description is given on how to add a new
model to the {\tt BSMPT} framework. A template class is provided to
simplify the implementation. For the definition of the tensors derived
from the Lagrangian in the gauge basis, we refer to the previous
version of the manual \cite{Basler:2018cwe} and focus here on the
differences between the two versions. In the
following, we give the steps to be performed to to modify the class template
to the desired model referred to as {\tt YourModel}. The concrete
implementation of a toy model will be given in \cref{sec:toyexample}.
\s

For the inclusion of a new model, the following steps
  have to be performed:
\begin{itemize}
\item [1)] The class template header file in 
\begin{center}
  \text{\home/include/BSMPT/models/ClassTemplate.h  }
\end{center}
 is copied to 
\begin{center}  
 \text{\home/include/BSMPT/models/YourModel.h}
\end{center}

 In this header file all occurrences of {\tt Class\_Template} have to be changed to {\tt YourModel}.

\item [2)] To call the correct model in the binaries the model ID has to be provided. {\tt YourModelID} has to be added as an enum in 
\begin{center}
  \text{\home/include/BSMPT/models/IncludeAllModels.h  }
\end{center}
below the already provided models {\it e.g.}~like 
\begin{lstlisting}[language=C]
enum class ModelIDs
{
  ...  
  YourModelID,
  ...
  // DO NOT EDIT the part below
  stop
};    
\end{lstlisting} 
Right below in {\tt ModelNames}, a new entry for the new model is to be added through 
\begin{lstlisting}[language=C]
const std::unordered_map<std::string,ModelIDs> ModelNames{
  ... 
  {"modelname" , ModelIDs::YourModelID},
  ...
};
\end{lstlisting}

Note that {\tt modelname} has to be in lowercase and will be used to
call the model in the binaries.  

  \item[3)] In the file 
\begin{center}
  \text{\home/src/models/IncludeAllModels.cpp}
\end{center}
  the header for the new model has to be
  added via the additional include 
\begin{lstlisting}[language=C]
#include <BSMPT/models/YourModel.h>
\end{lstlisting}
Afterwards the function {\tt FChoose} have to be
extended by the additional switch statement\\ 
\begin{lstlisting}[language=C]
std::unique_ptr<Class_Potential_Origin> FChoose(ModelIDs choice){
using namespace  Models;
switch (choice) {
... 
case ModelIDs::YourModelID :return std::make_unique<YourModel>();break;
...
}
}
\end{lstlisting}
\item[4)] The template 
\begin{center}
  \text{\home/src/models/ClassTemplate.cpp}
\end{center}
  has to be copied to 
\begin{center}
  \text{\home/src/models/YourModel.cpp}
\end{center}
  and changed accordingly. All occurrences of {\tt Class\_Template}
  have to be changed to {\tt YourModel}. We provide comments in the
  class templates indicating what else needs to be changed. The
  definition of the tensor structure can be found in
  \cite{Basler:2018cwe}. The following functions have to be adapted
  for the new model: 
      \begin{itemize}
        \item[] {\tt Class\_Template}
        \item[] {\tt ReadAndSet}
        \item[] {\tt addLegendCT}
        \item[] {\tt addLegendTemp}
        \item[] {\tt addLegendTripleCouplings}
        \item[] {\tt addLegendVEV}
        \item[] {\tt set\_gen}
        \item[] {\tt set\_CT\_Pot\_Par}
        \item[] {\tt write}
        \item[] {\tt TripleHiggsCouplings}
        \item[] {\tt calc\_CT}
        \item[] {\tt SetCurvatureArrays}
        \item[] {\tt CalculateDebyeSimplified}
        \item[] {\tt VTreeSimplified}
        \item[] {\tt VCounterSimplified}
        \item[] {\tt Debugging}\footnote{In fact,
                the function {\tt Debugging} is only used by the Test executable and
                is provided only for the user to perform some checks.}
      \end{itemize}
\item[5)] As last step the model has to be added to the {\tt cmake} file. In 
\begin{center}
  \text{\home/src/models/CMakeLists.txt}
\end{center}
the additional include has to be added
\begin{lstlisting}[language=C]
${header_path}/YourModel.h
\end{lstlisting}
and the file {\tt YourModel.cpp} has to be linked by adding 
\begin{lstlisting}[language=C]
set(src 
...
YourModel.cpp
... 
)
\end{lstlisting}
\end{itemize}
After performing these steps and rebuilding the program by following
the steps described in \cref{sec:Installation} {\tt YourModel} should
compile. The concrete implementation of the model parameters is
described in the following section by considering a simple toy
model. The corresponding tensors are defined in
\cite{Basler:2018cwe}. 

\subsection{Toy Model Example \label{sec:toyexample}}
As example we take a model with one scalar particle $\phi$ which
develops a VEV $v$, couples to one fermion $t$ with the Yukawa
coupling $y_t$, and to one gauge boson $A$ with the gauge coupling
$g$. The relevant pieces of the Lagrangian are given by  ($\Phi = \phi
  + v$)
\begin{subequations}  

\begin{align}
  -\mathcal{L}_S & = \frac{m^2}{2} \left(\phi+v\right)^2 +
  \frac{\lambda}{4!}
  \left(\phi+v\right)^4 \label{eq:templatetree}            \\
  -\mathcal{L}_F & = y_t t_L t_R \left(\phi + v\right)     \\
  \mathcal{L}_G  & = g^2A^2\left(\phi+v\right)^2 \,.
\end{align}
\end{subequations}
We therefore have $i,j,k,l = 1 , I,J = 1,2, a,b = 1$ for the tensors
defined in \cite{Basler:2018cwe}. Here $I,J=1,2$ corresponds to $t_L$ and
$t_R$, the left- and right-handed projections of the fermion $t$.
The tensors are given by
\begin{subequations}
\begin{align}
  L^i      & = \left.\partial_{v} \left(-\mathcal{L}_S \right)\right|_{\phi=0,v=0} = 0     \\
  L^{ij}   & = \left.\partial_{v}^2 \left(-\mathcal{L}_S \right)\right|_{\phi=0,v=0} = m^2 \\
  L^{ijk}  & =  \left.\partial_{v}^3 \left(-\mathcal{L}_S \right)\right|_{\phi=0,v=0} = 0  \\
  L^{ijkl} & =  \left.\partial_{v}^4 \left(-\mathcal{L}_S
  \right)\right|_{\phi=0,v=0} = \lambda                                                    \\
  Y^{IJk}  & = \begin{cases} 0   & I = J \; (I,J=t_L,t_R)   \\
    y_t & I \ne J \; (I,J=t_L,t_R)\end{cases}                                                  \\
  G^{abij} & = \partial_A^2 \partial_v^2 \left( \mathcal{L}_G \right) =
  4g^2 \;.
\end{align}
\end{subequations}
The counterterm potential reads

\begin{align}
  V^{\text{CT}} & = \frac{\delta m^2}{2} \left(\phi+v\right)^2 + \frac{\delta
    \lambda}{4!} \left(\phi+v\right)^4 + \delta T \left(\phi+
  v\right) \;. \label{eq:templatecounterpot}
\end{align}
The renormalisation conditions yield the following relations for the counterterms
\begin{subequations}
\begin{align}
  \delta T + v \delta m^2  + \frac{1}{6} v^3 \delta \lambda
                                           & = -\left.\partial_{\phi} V^{\text{CW}}\right|_{\phi=0} \\
  \delta m^2 + \frac{v^2}{2}\delta \lambda & = -\left. \partial^2_{\phi}
  V^{\text{CW}}\right|_{\phi=0} \;.
\end{align}
\end{subequations}
The resulting system of equations is overconstrained. The tadpole
counterterm is chosen to be arbitrary as
\begin{equation}
\delta T = t \;, \quad \mbox{with} \quad t \in \mathbb{R} \;,
\label{eq:cttad}
\end{equation}
yielding the remaining counterterms as 
\begin{subequations}
\begin{align}
\delta \lambda =& \frac{3t}{v^3} + \frac{3}{v^3} \left(\left.\partial_{\phi}
V^{\CW}\right|_{\phi=0}\right) - \frac{3}{v^2} \left(\left. \partial^2_{\phi}
V^{\CW}\right|_{\phi=0} \right) \label{eq:ctlambda} \\
\delta m^2 =& -\frac{3}{2v} \left( \left.\partial_{\phi}
V^{\CW}\right|_{\phi=0}\right) + \frac{1}{2} \left(\left. \partial^2_{\phi}
V^{\CW}\right|_{\phi=0}\right) - \frac{3t}{2v} \;. \label{eq:ctms}
\end{align}
\end{subequations}
In the following we will discuss the additional steps needed for the
implementation of this concrete toy model.  
 
\subsubsection{Proper Link of the Model}
As a first step we follow the steps described at the beginning of 
\cref{sec:newmodels}. The corresponding model ID {\tt ToyModel} and
model name {\tt toymodel} are added to {\tt IncludeAllModels.h} and
{\tt IncludeAllModels.cpp}, respectively. For the model header file
{\tt ToyModel.h} and the {\tt ToyModel.cpp} file the
template class
files can be used, where the occurrences of {\tt ClassTemplate} have
to be changed accordingly. When the steps 1-5 described at the
beginning of
\cref{sec:newmodels} are complete some additional information is
needed.  
In the header file {\tt ToyModel.h} the variables for the potential
and for the remaining Higgs coupling parameters as well as for the
counterterm constants have to
be added by including 
\begin{lstlisting}[language=C]
  double ms, lambda, dms, dlambda, dT, yt, g;
\end{lstlisting} 
Here {\tt ms} denotes the mass parameter squared, $m^2$, and {\tt dms},
{\tt dlambda} are the counterterms $\delta m^2$, $\delta
\lambda$. Note that the parameters do not need to be set yet,
they are only declared here.
\subsubsection{ToyModel.cpp}
In this section, we will briefly discuss which essential adaptions in
{\tt ToyModel.cpp} are needed to implement a working model into {\tt
  \ewBSMPT}. The user is further guided by comments in the template
source code.  
\paragraph*{Toy\_Model()}
The numbers of Higgs particles, potential parameters,
counterterms and VEVs have to be specified in the constructor
  {\tt Toy\_Model}() and the variable 'Model' has to be set to
the selected model. For the toy model example, this corresponds to 
\begin{lstlisting}[language=C]
Toy_Model::Toy_Model(){  
  Model = ModelID::ModelIDs::TOYMODEL; 
  NNeutralHiggs = 1; // Number of neutral Higgs bosons
  NChargedHiggs = 0; // Number of charged Higgs bosons
  nPar          = 2; // Number of independent input parameters
  nParCT        = 3; // Number of counterterms
  nVEV          = 1; // Number of non-zero VEVs
  NHiggs        = NNeutralHiggs+NChargedHiggs; 
  VevOrder.resize(nVEV);
  VevOrder[0]   = 0;  
  UseVTreeSimplified = false; // Option for using a simplified tree-level potential-> false uses the generic implementation
  UseVCounterSimplified = false; // Option for using a simplified counterterm potential-> false uses the generic implementation
}
\end{lstlisting}
It should be emphasized that internally the Higgs potential is built
by the vector of the Higgs field containing all fields, with a length
of {\tt NHiggs}. For the toy model this vector would have only one
component $\phi$. For a more complex model this might not be the  
case. However, not all field components need to have a non-zero
VEV. In this case, in order to determine which field components should be taken
into account in the minimisation of the potential, the vector {\tt
  VevOrder} has to specify which field components of the 
vector with length {\tt NHiggs} have a VEV. The code  
\begin{lstlisting}[language=C]
  VevOrder[0] = 0;
\end{lstlisting}
would attribute a VEV to the first
component of the field vector. For
instance, our implementation of the CxSM has the following constructor 
\begin{lstlisting}[language=C]
Class_CxSM::Class_CxSM()
{
  Model = ModelID::ModelIDs::CXSM; 
  NNeutralHiggs = 4; // Number of neutral Higgs bosons at T = 0
  NChargedHiggs=2; // Number of charged Higgs bosons  at T = 0 (all d.o.f.)
  nPar = 10+3; // Number of parameters in the tree-level Lagrangian
  nParCT = 9+6; // Number of parameters in the counterterm potential

  nVEV=3; // Number of VEVs to minimize the potential
  NHiggs = NNeutralHiggs+NChargedHiggs;

  VevOrder.resize(nVEV);
  VevOrder[0] = 3;
  VevOrder[1] = 4;
  VevOrder[2] = 5;

  UseVTreeSimplified = false;
  UseVCounterSimplified = false;
}
\end{lstlisting}
implying that the Higgs field vector has fields with
VEVs at the positions $3,4$ and $5$. The resulting three-dimensional
vector containing all field components with a VEV of the CxSM
corresponds then to the vacuum structure of the given model.

\paragraph*{Input Read-in}
{\tt \ewBSMPT} reads the input through {\tt std::stringstream} where
each parameter point is 
given as a simple line with tab separation. The user has to specify
the order of the respective input parameters in the function {\tt
  ReadAndSet()}. For the toy model under consideration two input
parameters, $m^2$ and $\lambda$, have to be specified 
\begin{lstlisting}[language=C]
void Toy_Model::ReadAndSet(const std::string& linestr, std::vector<double>& par )
{
  std::stringstream ss(linestr);
  double tmp;
  double lms,llambda;
  for(int k=1;k<=2;k++)
    {
      ss>>tmp;
      if(k==1) lms = tmp; // Class member lms is set to the respective input parameter
      else if(k==2) llambda = tmp; //Class member llambda is set to the respective input parameter
    }
  par[0] = lms; 
  par[1] = llambda;
  set_gen(par); // Setting of all model parameters with the given input parameters par
  return ;
}
\end{lstlisting}
where {\tt par} contains all input parameters in a fixed order. The
function {\tt set\_gen} sets all remaining model parameters
in accordance with the input parameters. This function has changed with respect to the previous {\tt BSMPT}
  version and old files have to be adapted for compatibility with {\tt
  BSMPT v2}.

\paragraph*{set\_gen()}
This function allows to express all model parameters in accordance with
the input parameters. This function is called in the {\tt
  ReadAndSet()} function after the input parameter vector {\tt par} is
set. Further SM input parameters are provided by  
\begin{center}
  \text{\home/include/BSMPT/models/SMparam.h}
\end{center}
and can be used to set up the parameters of the toy model. For the toy
model, the {\tt set\_gen} function would look like  
\begin{lstlisting}[language=C]
void Toy_Model::set_gen(const std::vector<double>& par) {
  ms = par[0]; //Class member is set in accordance with the input parameters
  lambda = par[1]; //Class member is set in accordance with to the input parameters
  g=C_g; //SM SU(2) gauge coupling  --> SMparam.h
  yt = std::sqrt(2)/C_vev0 * C_MassTop; //Top Yukawa coupling --> SMparam.h
  scale = C_vev0; //Renormalisation scale is set to the SM VEV
  vevTreeMin.resize(nVEV);
  vevTree.resize(NHiggs); 
  vevTreeMin[0] = C_vev0; 
  vevTree=MinimizeOrderVEV(vevTreeMin);
  if(!SetCurvatureDone) SetCurvatureArrays();
}
\end{lstlisting}
More complex models might require to
set further parameters of the Lagrangian. The analytic formulas expressing
these parameters in terms of the input parameters can be implemented in
{\tt set\_gen}.  

\paragraph*{SetCurvatureArrays()}
The tensors of the Lagrangian of the new model have to be implemented
in the function {\tt SetCurvatureArrays()}. The
notation follows the definition in \cite{Basler:2018cwe} and the corresponding variable names are given by
\begin{subequations}
  \begin{align}
    \text{Curvature\_Higgs\_L1}[i]            & \overset{\wedge}{=} L^i        \\
    \text{Curvature\_Higgs\_L2}[i][j]         & \overset{\wedge}{=} L^{ij}     \\
    \text{Curvature\_Higgs\_L3}[i][j][k]      & \overset{\wedge}{=} L^{ijk}    \\
    \text{Curvature\_Higgs\_L4}[i][j][k][l]   & \overset{\wedge}{=} L^{ijkl}   \\
    \text{Curvature\_Gauge\_G2H2}[a][b][i][j] & \overset{\wedge}{=} G^{abij}   \\
    \text{Curvature\_Quark\_F2H1}[I][J][k]    & \overset{\wedge}{=} Y^{IJk}\,.
  \end{align}
\end{subequations}
Technically, it is possible to use {\tt Curvature\_Quark\_F2H1} to store all quarks
and leptons simultaneously, but as they do not mix the program provides besides
{\tt Curvature\_Quark\_F2H1} where $I,J$ run over all quarks, also the
structure {\tt Curvature\_Lepton\_F2H1[I][J][k]} where $I,J$ run over all
leptons. Only non-zero tensor coefficients need to be provided. The
implementation for the toy model would look like 
\begin{lstlisting}[language=C]
void Toy_Model::SetCurvatureArrays(){
  initVectors();
  SetCurvatureDone=true;
  for(size_t i=0;i<NHiggs;i++) HiggsVev[i] = vevTree[i];
  Curvature_Higgs_L2[0][0] = ms;
  Curvature_Higgs_L4[0][0][0][0] = lambda;
  Curvature_Gauge_G2H2[0][0][0][0] = 4*std::pow(g,2);
  Curvature_Quark_F2H1[1][0][0] = yt;
  Curvature_Quark_F2H1[0][1][0] = yt;
}
\end{lstlisting}

\paragraph*{set\_CT\_Pot\_Par()}
Analogously, the curvature terms for the counterterm potential have to
be set. The corresponding tensors are defined as in the tree-level
potential with the additional suffix $\_CT\_$. For the toy model the
implementation would look like 
\begin{lstlisting}[language=C]
void Toy_Model::seT_cT_Pot_Par(const std::vector<double>& par){

	dT = par[0];
	dlambda = par[1];
	dms = par[2];

	Curvature_Higgs_CT_L1[0] = dT;
	Curvature_Higgs_CT_L2[0][0] = dms;
	Curvature_Higgs_CT_L4[0][0][0][0] = dlambda;
}

\end{lstlisting}

\paragraph*{calc\_CT()}
The counterterms are computed numerically in the function  {\tt
    calc\_CT()}. To do so, the user has to implement the
formulae for the counterterms that were derived beforehand
analytically in terms of the derivatives of the Coleman-Weinberg
potential, {\it cf.}~Eqs.~(\ref{eq:cttad}), (\ref{eq:ctlambda}) and
(\ref{eq:ctms}) for our template
model. The derivatives of $V^{\CW}$ are provided by the program
through the function calls {\tt WeinbergFirstDerivative} and {\tt
    WeinbergSecondDerivative}. 
\begin{lstlisting}[language=C]
std::vector<double> Toy_Model::calc_CT() const {
  std::vector<double> parCT;
  ...
  std::vector<double> WeinbergNabla,WeinbergHesse;
  WeinbergNabla = WeinbergFirstDerivative(); //Call for the calculation of the first derivative of the Coleman-Weinberg potential
  WeinbergHesse = WeinbergSecondDerivative();//Call for the calculation of the second derivative of the Coleman-Weinberg potential
  ...
  double t = 0;
  parCT.push_back(t); // Tadpole counterterm dT
  parCT.push_back(3.0*t/std::pow(C_vev0,3) + 3.0/std::pow(C_vev0,3) *
  NablaWeinberg(0) -3.0/std::pow(C_vev0,2) *HesseWeinberg(0,0)); //  Analytic formula of the counterterm dlambda
  parCT.push_back(-3.0/(2*std::pow(C_vev0,2)) *NablaWeinberg(0) + 1.0/2.0 *HesseWeinberg(0,0) -3.0*t/(2*C_vev0)); // Analytic formula of the counterterm dms
  return parCT;
}
\end{lstlisting}
The remaining code base (indicated by '\dots') can be copied from
the template class file.     

\paragraph*{TripleHiggsCouplings()}
This function provides the trilinear loop-corrected
Higgs self-couplings as obtained from the effective potential. They
are calculated from the third derivative of the Higgs potential
with respect to the Higgs fields in the gauge basis and then rotated
to the mass basis. 
The scalar mass matrix is built from all the scalar degrees of freedom of the theory, which we refer to as the gauge basis \cite{Basler:2018cwe}.
Since the numerical routines in \texttt{Eigen} return the mass eigenvalues ordered by ascending masses, the corresponding eigenstates have to identified with the respective Higgs bosons.   
Therefore, it is necessary to map both bases, the mass ordered mass eigenstate basis and the physical Higgs boson basis. The mapping is achieved through the vector {\tt HiggsOrder(NHiggs)} \textit{e.g.}
\begin{lstlisting}[language=C]
for(size_t i=0;i<NHiggs;i++) HiggsOrder[i]=value[i];
\end{lstlisting}
The mapping {\tt value} is defined by the user according
to the ordering that this desired in the Higgs basis. Thus {\tt
    HiggsOrder}$[0]=5$ {\it e.g.}~would assign the 6th lightest particle to the first
position. The particles can be selected through
the mixing matrix elements.\\ 
All following functions add specific legends to the output header. 

\paragraph*{addLegendTripleCouplings()}
 The function {\tt addLegendTripleCouplings} extends the legend by the column names for
the trilinear Higgs couplings derived from the tree-level, the
counterterm and the Coleman-Weinberg potential. In order to do so,
the user first has to make sure to define the names of the Higgs
particles of the model in the vector {\tt particles}. In the toy model
only one Higgs boson is present, which we refer to as $H$ and hence set 
\begin{lstlisting}[language=C]
  particles[0] = "H";
\end{lstlisting}
\paragraph*{addLegendTemp()}
In this function, the column names for $T_c$, $v_c$ and the VEVs are added to the
legend. The order should be $T_c$,  $v_c$ and then the names of the individual
VEVs. These VEVs have to be added in the same order as given
in the function {\tt MinimizeOrderVEV}.

\paragraph*{addLegendVEV()}
This function adds the column names for the VEVs
that are given out. The order has to be the same as given in the
function {\tt MinimizeOrderVEV}.

\paragraph*{addLegendCT()}
In this function, the legend for the counterterms is added. The order of
the counterterms has to be same as the one set in the function {\tt
    set\_CT\_Pot\_Par(par)}.

\paragraph*{VTreeSimplified, VCounterSimplified}
The functions  
\begin{lstlisting}[language=C]
  VTreeSimplified(const std::vector<double>& v)
  VCounterSimplified(const std::vector<double>& v)
\end{lstlisting}
can be used to
explicitly implement the formulas for the tree-level and counterterm potential in
terms of the classical fields $\omega$, in our
example these are Eqs.~(\ref{eq:templatetree}) and
Eq.~(\ref{eq:templatecounterpot}), respectively, with
$\phi=0$ and $v \equiv \omega$. Implementing these
may improve the runtime of the programs. An example is given in the template class.

\paragraph*{write()}
The function {\tt write()} allows us to print all needed parameters in
the terminal and might be used for the debugging
procedure or some investigation of parameter behaviour. One
minimalist example for the toy model could look like  
\begin{lstlisting}[language=C]
void Class_Template::write() const {

  std::cout << "The parameters are : " << std::endl;
  std::cout << "lambda = "  << lambda << std::endl
            << "m^2 = " << ms     << std::endl;

  std::cout << "The counterterm parameters are : " << std::endl;
  std::cout << "dT = "      << dT       << std::endl
            << "dlambda = " << dlambda  << std::endl
            << "dm^2 = "    << dms      << std::endl;
  std::cout << "Some interesting number" << lambda * ms << std::endl;
}
  
\end{lstlisting}

\section{Using BSMPT as a library \label{sec:lib}}
During the installation, {\tt \ewBSMPT} is exported as a {\tt cmake}
package. It is possible to link {\tt \ewBSMPT } in an existing {\tt
  cmake} project by including the line   
\begin{lstlisting}[language=C]
find_package(BSMPT 2.0)
\end{lstlisting}
which provides the {\tt BSMPT::Models}, {\tt BSMPT::Minimizer},
{\tt BSMPT::ThermalFunctions} and \\ {\tt BSMPT::Utility}
modules. Furthermore, the modules {\tt BSMPT::Baryo}, {\tt
  BSMPT::LibWallThickness} and {\tt BSMPT::Kfactors} are available if the {\tt boost} library is provided.

\section{Upgrading from BSMPT v1.x to v2 \label{sec:upgrade}}
With the upgrade from v1 to v2 the interface to the basic
functions and most of the return types of the functions have
changed. Private implementations for {\tt BSMPT} have to be adapted to
be compatible with {\tt\ewBSMPT}. It is recommended to use the
template class for new model implementations. If there are remaining
compatibility problems, contact us either through github or via mail 
\url{bsmpt@lists.kit.edu}. 

\section{Conclusions \label{sec:concl}}
We have presented the {\tt C++} code {\tt BSMPT v2}, an extension of
the previous code {\tt BSMPT} that calculates the strength of the
electroweak phase transition in Higgs sector extensions beyond the
SM. The main new feature is the 
implementation of the calculation of the BAU in two different
approximations, the \FH and the \VIA approach. Presently, this is
applicable for the C2HDM and will be extended to include further
models in future upgrades of the code. Furthermore, a new model, the
CxSM has been included. We also added the possibility
to change the renormalisation scale in the computation of the
loop-corrected effective potential. The newly implemented model CxSM,
the computation of the BAU in the two approaches \FH and \VIA and the
additional changes and new features have been presented in detail
along with a description of the procedure for the implementation of new
models. This has been changed as well with respect to the previous
version of {\tt BSMPT}. The code can be downloaded at the url: \download. 


\section{Acknowledgments}
The research of MM was supported by the Deutsche Forschungsgemeinschaft (DFG, German
Research Foundation) under grant 396021762 - TRR
257. P.B. acknowledges financial support 
by the Graduiertenkolleg GRK 1694: Elementarteilchenphysik bei h\"ochster Energie und h\"ochster
Pr\"azision. 
JM acknowledges support by the BMBF-Project 05H18VKCC1. 
We want to thank Jonas Wittbrodt for help with the {\tt cmake}
configuration. We thank Stephan Huber for helpful
discussions.

\section*{Appendix}
\appendix
\section{Tadpole Equations for the CxSM \label{app:tadpole}}
The local minima of the tree-level potential of \cref{eq:CxSMPot} are
given by 
\begin{align}
	0 =& \frac{\partial V}{\partial \Phi^\dagger} \\
	0 =& \frac{\partial V}{\partial \mathbb{S}^\dagger} \,.
\end{align}

This yields the following three equations 
\begin{subequations}
\begin{align}
	0 =& \frac{v}{4} \left( v^2 \lambda + \delta_2 v_a^2 + \delta_2 v_s^2 + 2m^2\right) \\
	0 =& \mbox{Re}(a_1) \sqrt{2} - \frac{\mbox{Im}(b_1)}{2} v_a + \frac{v_s}{4} \left( d_2 v_a^2 + d_2 v_s^2 + \delta_2 v^2 + 2\mbox{Re}(b_1) + 2 b_2\right) \\
	\mbox{Im}(a_1) \sqrt{2} + \mbox{Im}(b_1) \frac{v_s}{2} =& \frac{v_a}{4} \left( d_2 v_a^2 + d_2 v_s^2 + \delta_2 v^2 - 2\mbox{Re}(b_1) + 2 b_2\right) \,.
\end{align}
\end{subequations}

For the different vacuum configurations they are solved for different
parameters. The solutions are

\paragraph*{Case $v_s \neq 0 \wedge v_a \neq 0$}
\begin{subequations}
\begin{align}
\mbox{Re}(a_1) =& \frac{\sqrt{2}d_2}{4} v_s \left( v_s^2 +v_a^2 \right) - \frac{\sqrt{2} \delta_2}{4} v_s v^2 + \mbox{Im}(a_1) \frac{v_s}{v_a} + \frac{\sqrt{2} \mbox{Im}(b_1)}{4} \left( v_a + \frac{v_s^2}{v_a} \right) - \frac{b_2 v_s}{\sqrt{2}} \\
\mbox{Re}(b_1) =& b_2 + \frac{d_2}{2} \left( v_s^2 + v_a^2 \right) + \frac{\delta_2}{2} v^2 - \frac{2\sqrt{2} \mbox{Im}(a_1)}{v_a} - \mbox{Im}(b_1) \frac{v_s}{v_a} \\
m^2 =& - \frac{\delta_2}{2} \left( v_s^2 + v_a^2 \right) - \frac{\lambda}{2} v^2 \,.
\end{align}
\end{subequations}

\paragraph*{Case $v_s \neq 0 \wedge v_a = 0$}
\begin{subequations}
\begin{align}
\mbox{Im}(a_1) =& - \frac{\sqrt{2}}{4} \mbox{Im}(b_1) v_s \\
\mbox{Re}(b_1) =& -b_2 - \frac{d_2 v_s^2}{2} - \frac{\delta_2 v^2}{2} - \frac{2\sqrt{2} \mbox{Re}(a_1)}{v_s} \\
m^2 =& - \frac{\delta_2 v_s^2}{2} - \frac{\lambda v^2}{2} \,.
\end{align}
\end{subequations}

\paragraph*{Case $v_s = 0 \wedge v_a \neq 0$}
\begin{subequations}
\begin{align}
\mbox{Re}(a_1) =& \frac{\sqrt{2}}{4} \mbox{Im}(b_1) v_a \\
\mbox{Re}(b_1) =& b_2 + \frac{d_2 v_a^2}{2} + \frac{\delta_2 v^2}{2} - \frac{2\sqrt{2} \mbox{Im}(a_1)}{v_a} \\
m^2 =& - \frac{\delta_2 v_a^2}{2} - \frac{\lambda v^2}{2} \,.
\end{align}
\end{subequations}

\paragraph*{Case $v_s = 0 \wedge v_a = 0$}
\begin{subequations}
\begin{align}
\mbox{Im}(a_1) =& 0 \\
\mbox{Re}(a_1) =& 0 \\
m^2 =& -\lambda \frac{v^2}{2} \,.
\end{align}
\end{subequations}


\end{document}